\documentclass[journal]{IEEEtran}

\IEEEoverridecommandlockouts
\usepackage{pifont}
\usepackage{booktabs,siunitx}
\usepackage{multirow}
\usepackage{cite}
\usepackage{amsmath,amssymb,amsfonts}
\usepackage{algorithm}
\usepackage{algorithmic}
\usepackage{graphicx}
\usepackage{textcomp}
\usepackage{xcolor}
\usepackage{optidef} 
\usepackage{verbatim}
\usepackage{todonotes}                                                                          
\usepackage{caption}
\expandafter\def\expandafter\normalsize\expandafter{%
    \normalsize%
    \setlength\abovedisplayskip{0pt}%
    \setlength\belowdisplayskip{8pt}%
    \setlength\abovedisplayshortskip{-1pt}%
    \setlength\belowdisplayshortskip{2pt}%
}
\usepackage{subcaption}
\usepackage{todonotes}
\usepackage{titlesec}         
\usepackage[toc,title,page]{appendix}
\DeclareMathSizes{10}{9}{6}{5}
\DeclareMathSizes{10}{9}{6}{5}

\begin{document}
\title{\LARGE{Resource-Efficient Beam Prediction in mmWave Communications\\with Multimodal Realistic Simulation Framework}}
\author{Yu~Min~Park,~\IEEEmembership{Member,~IEEE},~Yan~Kyaw~Tun~\IEEEmembership{Member,~IEEE},~Eui-Nam~Huh,~\IEEEmembership{Member,~IEEE}\\~Walid~Saad,~\IEEEmembership{Fellow,~IEEE},~and~Choong~Seon~Hong,~\IEEEmembership{Fellow,~IEEE} \vspace{-0.2in}

\IEEEcompsocitemizethanks{ \IEEEcompsocthanksitem Yu Min Park is with the Department of Computer Science and Engineering, Kyung Hee University, Yongin-si, Gyeonggi-do 17104, Rep. of Korea, and also with the Bradley Department of Electrical and Computer Engineering, Virginia Tech, Alexandria, VA, 22305, USA, email: yumin0906@khu.ac.kr.}
\IEEEcompsocitemizethanks{ \IEEEcompsocthanksitem Yan Kyaw Tun is with the Department of Electronic Systems, Aalborg University, A. C. Meyers Vænge 15, 2450 København, email: ykt@es.aau.dk.}
\IEEEcompsocitemizethanks{ \IEEEcompsocthanksitem Walid Saad is with the Bradley Department of Electrical and Computer Engineering, Virginia Tech, Alexandria, VA, 22305, USA, email: walids@vt.edu.}
\IEEEcompsocitemizethanks{ \IEEEcompsocthanksitem Eui-Nam Huh and Choong Seon Hong are with the Department of Computer Science and Engineering, Kyung Hee University, Yongin-si, Gyeonggi-do 17104, Rep. of Korea, emails: johnhuh@khu.ac.kr, cshong@khu.ac.kr.}
}

\markboth{Journal of \LaTeX\ Class Files,~Vol.~14, No.~8, August~2015}
{Shell 
\MakeLowercase{\textit{et al.}}: Bare Demo of IEEEtran.cls for IEEE Journals}
\maketitle
\begin{abstract}
Beamforming is a key technology in millimeter-wave (mmWave) communications that improves signal transmission by optimizing directionality and intensity. However, conventional channel estimation methods, such as pilot signals or beam sweeping, often fail to adapt to rapidly changing communication environments. To address this limitation, multimodal sensing-aided beam prediction has gained significant attention, using various sensing data from devices such as LiDAR, radar, GPS, and RGB images to predict user locations or network conditions. Despite its promising potential, the adoption of multimodal sensing-aided beam prediction is hindered by high computational complexity, high costs, and limited datasets. \textcolor{black}{Thus, in this paper, a novel resource-efficient learning framework is introduced for beam prediction, which leverages a custom-designed cross-modal relational knowledge distillation (CRKD) algorithm specifically tailored for beam prediction tasks, to transfer knowledge from a multimodal network to a radar-only student model, achieving high accuracy with reduced computational cost.} To enable multimodal learning with realistic data, a novel multimodal simulation framework is developed while integrating sensor data generated from the autonomous driving simulator CARLA with MATLAB-based mmWave channel modeling, and reflecting real-world conditions. The proposed CRKD achieves its objective by distilling relational information across different feature spaces, which enhances beam prediction performance without relying on expensive sensor data. Simulation results demonstrate that CRKD efficiently distills multimodal knowledge, allowing a radar-only model to achieve $94.62\%$ of the teacher performance. In particular, this is achieved with just $10\%$ of the teacher network’s parameters, thereby significantly reducing computational complexity and dependence on multimodal sensor data.
\end{abstract}
\begin{IEEEkeywords}
Multimodal learning, simulation framework, beamforming, sensing-aided beam prediction, relational knowledge distillation, cross-modal learning.
\end{IEEEkeywords}

\IEEEpeerreviewmaketitle

\section{Introduction and Background}
\label{introduction}
Millimeter-wave (mmWave) communications are widely recognized as a key enabler for next-generation wireless systems, because of their ability to provide high data rates and support numerous bandwidth-intensive applications \cite{tataria20216g}. A cornerstone technology in mmWave communication is beamforming, which directs wireless signals to specific spatial directions to improve signal strength and transmission quality \cite{brilhante2023literature}. However, mmWave signals experience high path loss and narrow beamwidth. This, in turn, makes it challenging to perform accurate beam alignment to maintain reliable links, particularly in dynamic or high mobility scenarios. A commonly employed solution for beam alignment is beam sweeping, in which the transmitter (Tx) and receiver (Rx) systematically scan multiple beam directions to find the optimal alignment. Although this approach is straightforward to implement, it can be inefficient in rapidly changing environments, introducing considerable overhead in terms of time and energy \cite{li2020beam}.

One promising approach for overcoming these limitations is to leverage multimodal sensing-aided beam prediction using data from sensors such as LiDAR, radar, GPS, and RGB cameras to monitor user trajectories and network conditions \cite{wen2024survey}. By incorporating this information, multimodal systems can forego or accelerate sequential beam scanning, allowing quicker and more flexible beam alignment with reduced overhead \cite{salehi2022deep}. This technique is well-suited for applications that require low latency, such as autonomous driving and drone communications \cite{xiao2021survey}. As such, multimodal sensing-aided beam prediction is becoming an important technology for future wireless networks \cite{khan2023machine}. However, designing practical multimodal sensing-aided beam prediction approaches requires overcoming a number of key challenges. First, deploying high-resolution sensors such as LiDAR or high-frame rate cameras at every base station is costly, and extensive use of cameras also raises privacy concerns \cite{jiang2022lidar}. Second, large-scale transformer-based fusion models are often needed to effectively leverage sensor data, but they can be computationally heavy \cite{wu2022blockage, cheng2023intelligent}. Third, creating comprehensive multimodal datasets is non-trivial, as existing public datasets often lack certain sensing modalities or have environment-specific constraints.

\begin{table*}[t]
    \centering
    \caption{Typical measurement and simulation public datasets for multimodal wireless communication.}
    \begin{tabular}{lcccccccccc}
        \hline
        \textbf{Dataset} 
        & \multicolumn{4}{c}{\textbf{Sensory data}} 
        & \multicolumn{2}{c}{\textbf{Communication data}} 
        & \textbf{Weather} 
        & \textbf{Multi-Scenario} 
        & \textbf{Source} \\
        \cmidrule(lr){2-5}\cmidrule(lr){6-7}
        & {RGB} 
        & {Depth map} 
        & {LiDAR} 
        & {Radar} 
        & {mmWave} 
        & {Massive MIMO} 
        & {Sunny, rainy, snowy} 
        & 
        & \\
        \hline
        DeepSense\,6G \cite{alkhateeb2023deepsense} & \ding{51} & \ding{55} & \ding{51} & \ding{51} & \ding{51} & \ding{51} & \ding{51} & \ding{51} & Measurement \\
        WLADO \cite{waldo_dataset} & \ding{55} & \ding{55} & \ding{55} & \ding{55} & \ding{55} & \ding{55} & \ding{55} & \ding{55} & Measurement \\
        Vi-Fi \cite{liu2022vi} & \ding{51} & \ding{51} & \ding{55} & \ding{55} & \ding{55} & \ding{55} & \ding{55} & \ding{55} & Measurement \\
        NEU \cite{salehi2022deep} & \ding{51} & \ding{55} & \ding{51} & \ding{55} & \ding{51} & \ding{51} & \ding{55} & \ding{51} & Measurement \\
        DeepMIMO \cite{Alkhateeb2019} & \ding{51} & \ding{51} & \ding{55} & \ding{51} & \ding{51} & \ding{55} & \ding{55} & \ding{55} & Simulation \\
        LASSE \cite{klautau20185g} & \ding{51} & \ding{51} & \ding{51} & \ding{55} & \ding{51} & \ding{51} & \ding{55} & \ding{51} & Simulation \\
        ViWi \cite{Alrabeiah19} & \ding{51} & \ding{51} & \ding{51} & \ding{55} & \ding{51} & \ding{51} & \ding{55} & \ding{51} & Simulation \\
        V2X-Sim \cite{li2022v2x} & \ding{51} & \ding{51} & \ding{51} & \ding{55} & \ding{55} & \ding{55} & \ding{55} & \ding{55} & Simulation \\
        e-Flash \cite{steinmetzer2017compressive} & \ding{51} & \ding{55} & \ding{51} & \ding{55} & \ding{51} & \ding{55} & \ding{55} & \ding{55} & Simulation \\
        \hline
    \end{tabular}
    \label{tab:datasets}
\vspace{-.175in}
\end{table*}

\vspace{-.175in}
\subsection{Prior Works}
There has been a number of works that attempted to address the aforementioned challenges \cite{kutty2015beamforming, heng2021six, que2023joint, zhang2017codebook, alrabeiah2022neural, bian20243, nie2023vision, cui2024sensing, tariq2024deep, tian2023multimodal, charan2022vision, zhu2025advancing, alkhateeb2023deepsense, waldo_dataset, liu2022vi, Alkhateeb2019, klautau20185g,Alrabeiah19,li2022v2x,steinmetzer2017compressive, cazzella2024multi, cheng2023m, gharsallah2024mvx}, as detailed next.
\label{prior_work}
\subsubsection{Traditional Approaches for mmWave Beamforming:}
Traditional beamforming in mmWave systems has largely relied on exhaustive beam sweeping or heuristic codebook-based methods~\cite{kutty2015beamforming}. Beam sweeping is the most established beam optimization technique in mmWave communications, in which the transmitter and receiver systematically scan multiple beam directions to determine the one that provides the highest signal strength \cite{heng2021six}. In \cite{que2023joint}, the authors leveraged simultaneous localization and mapping (SLAM) techniques to significantly improve beam tracking and management in mmWave systems by incorporating geometric consistency and environmental features. Although straightforward to implement and effective in static or low-mobility environments, beam sweeping can lead to considerable overhead and latency, which becomes problematic in high-speed or rapidly varying channels where real-time adaptability is essential \cite{xiao2021survey}. To mitigate this issue, heuristic methods have also been explored, such as codebook-based beam selection, where candidate beams are rapidly identified, and greedy approaches that pick the beam delivering the highest instantaneous signal strength \cite{zhang2017codebook}. Although relatively simple and intuitive, these methods do not guarantee a global optimum and often fail in highly dynamic environments \cite{xiao2021survey}. Consequently, relying solely on beam sweeping or heuristic approaches to overcome frequent blockages and severe path loss in mmWave systems can lead to increased communication delays and increased energy consumption, ultimately hampering the reliability of networks that demand rapid and continuous beam adaptation \cite{alrabeiah2022neural}.

\subsubsection{Multimodal Sensing-aided Beam Prediction:}
To address the challenges of mmWave beamforming, recent studies propose to take advantage of sensor modalities such as LiDAR, radar, GPS, and RGB cameras for more accurate beamforming \cite{bian20243, nie2023vision, cui2024sensing, tariq2024deep, tian2023multimodal, charan2022vision, zhu2025advancing}. The works in \cite{nie2023vision, bian20243, cui2024sensing} used multimodal learning approaches that integrate LiDAR, radar, RGB, and GPS data to improve beam prediction accuracy. In \cite{tariq2024deep}, the authors proposed a deep quantum transformer network that fuses multimodal sensing data for robust mmWave beam prediction in integrated sensing and communication systems, demonstrating notable performance gains in real-world V2I scenarios. However, practical deployment of the solutions in \cite{nie2023vision, cui2024sensing, tariq2024deep} is hindered by sensor cost, increased computational complexity, and scalability concerns. Many existing models \cite{tian2023multimodal, charan2022vision, zhu2025advancing} for beam prediction assume the availability of various multimodal data, which may not be feasible in real-world deployments where the infrastructure is restricted to limited sensor modalities. In particular, current LiDAR systems are often prohibitively expensive, and the widespread deployment of cameras raises significant privacy concerns. Moreover, while transformer architectures like the ones used in \cite{nie2023vision, cui2024sensing, tariq2024deep, tian2023multimodal} can handle numerous multimodal tasks, they typically require significant computational resources to achieve fast inference. Without such hardware, real-world base stations cannot integrate such transformer-based solutions for beamforming purposes.

\subsubsection{Multimodal Datasets and Simulation Frameworks:}
To effectively train multimodal beam prediction models, comprehensive datasets containing multiple sensor modalities along with corresponding beamforming information are required. Table~\ref{tab:datasets} shows publicly available datasets for multimodal learning in wireless communications. Moreover, these datasets can be broadly categorized into real-world and virtual environment datasets, each with its own significant limitations. In particular, publicly available datasets often lack the comprehensive multimodal data necessary for effective learning and evaluation. Among real-world datasets \cite{alkhateeb2023deepsense, waldo_dataset, liu2022vi, salehi2022deep}, very few comprehensively include LiDAR, radar, and RGB data together. Hence, such approaches have very limited applicability in multimodal learning and sensor fusion research. Furthermore, most datasets are designed with a specific research focus, limiting their adaptability to explore new communication paradigms beyond their original scope. In addition, virtual environment datasets \cite{Alkhateeb2019, klautau20185g,Alrabeiah19,li2022v2x,steinmetzer2017compressive}, often generated using network simulation tools such as MATLAB or Sionna, differ significantly from real-world datasets in terms of environmental complexity, sensor characteristics, and noise modeling. These discrepancies reduce the generalizability of models trained solely on synthetic data, making them less effective in practical deployment scenarios. 

\begin{figure}[t]
    \centering
    \includegraphics[width=\columnwidth]{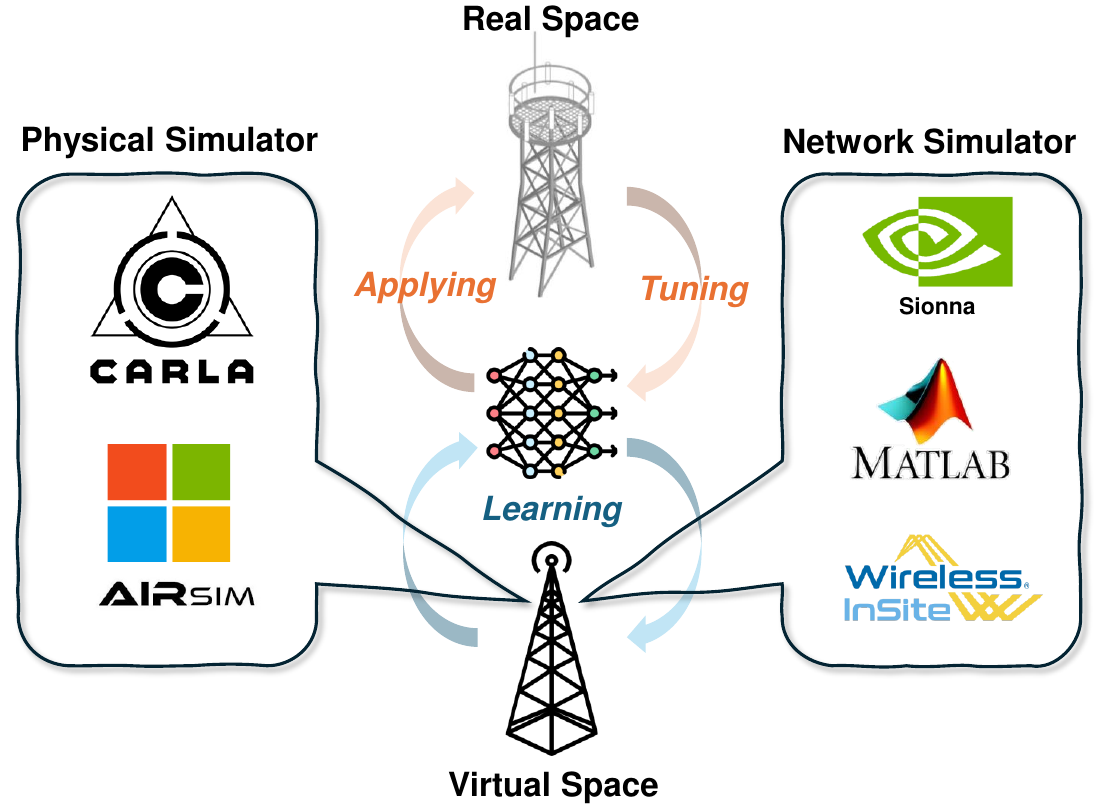}
    \caption{Model training in a virtual environment that combines multiple simulators.}
    \label{fig:digitaltwin}
\vspace{-.175in}
\end{figure}

To overcome these challenges, recent studies \cite{cazzella2024multi, cheng2023m, gharsallah2024mvx} proposed a realistic multimodal simulation framework that closely replicates real-world conditions in a virtual environment by integrating sensor data with beamforming information. This framework bridges the gap between synthetic and real-world datasets, enabling more robust and generalizable multimodal learning models for beam prediction in next-generation communication systems. To enhance the accuracy and realism of the generated data, the realistic multimodal simulation framework integrates an autonomous driving simulation platform with network simulation tools, as illustrated in Fig.~\ref{fig:digitaltwin}. The authors in \cite{cazzella2024multi} proposed a multimodal simulation framework for digital twin (DT) enabled vehicle-to-everything (V2X) communications, using CARLA for realistic sensor data generation and Remcom Wireless InSite for precise ray-tracing-based wireless channel modeling, demonstrated through a blockage handover task for V2X link restoration. Similarly, the work in \cite{cheng2023m} introduced M3SC, a comprehensive multimodal sensing-communication dataset, generated using AirSim, WaveFarer, and Wireless InSite, effectively aligning the physical space (LiDAR, RGB, radar) with the electromagnetic space (mmWave, channel impulse response (CIR) matrices) under various weather conditions and frequency bands. Furthermore, in \cite{gharsallah2024mvx}, the authors developed MVX-ViT, a co-simulation framework that integrates CARLA and Sionna to generate a multimodal V2X dataset, enabling AI-driven antenna position optimization. However, prior work has largely focused on perception tasks and lacks comprehensive experimentation on beamforming communication. To address this limitation, we propose a realistic multimodal simulation framework that combines CARLA with MATLAB, enabling detailed and diverse experiments on multimodal sensing and mmWave beamforming communication.

\subsubsection{Knowledge Distillation for Resource-Efficient Learning:}
Knowledge distillation (KD) compresses a high-capacity teacher model into a lightweight student model by training the student to mimic the teacher's output, thus preserving performance while significantly reducing computational requirements \cite{hinton2015distilling}. This approach is particularly beneficial in resource-constrained environments, such as edge devices or mobile platforms, where large models are impractical due to limited memory and power budgets. In standard KD, the teacher’s probabilistic outputs (soft labels) guide the student, allowing it to learn richer data distributions than is typically possible with hard labels alone. Beyond logit-based distillation, other KD variants include feature-based distillation, which transfers intermediate representations, and relation-based distillation, which preserves the structural relationships between data points in the latent space of the teacher \cite{gou2021knowledge}. Although most KD methods assume that both the teacher and the student share the same input modality, cross-modal knowledge distillation (CKD) extends KD to allow knowledge transfer between models trained in different sensor modalities \cite{yang2023categories}. In \cite{zhao2024crkd}, the authors proposed a teacher model trained on LiDAR and RGB data that can transfer its learned representations to a student model using radar and RGB input, thereby reducing the reliance on computationally expensive sensors during inference. Similarly, other CKD approaches \cite{huo2024c2kd} and \cite{ni2022cross} have been applied primarily to perception tasks such as object detection or classification. Although these works \cite{zhao2024crkd, huo2024c2kd, ni2022cross} demonstrate the potential of CKD for efficient inference, it does not explore or address beamforming communication. Thus, its applicability to real-time, resource-constrained beam prediction in 6G networks remains limited. To address these challenges, we investigate a radar-only student model for beam prediction that benefits from cross-modal relational knowledge distilled from transformer based multimodal teacher.

\vspace{-.175in}
\subsection{Contributions}
\label{contributions}
The main contribution of this paper is a novel resource-efficient learning approach based on Cross-modal Relational Knowledge Distillation (CRKD) for optimal beam prediction in a multi-vehicle-to-infrastructure (V2I) environment. We first define the beam prediction problem, which aims to maximize the received signal strength (RSS) between multiple vehicles. To generate the necessary training data, we introduce a realistic multimodal simulation framework that integrates traditional communication tools (MATLAB) with autonomous driving simulators (CARLA). This framework enables diverse multimodal experiments using realistic sensor data. Using this multimodal dataset, we propose a CRKD-based approach for efficient radar-only beam prediction. Specifically, our method transfers knowledge from a teacher network trained in multiple sensor modalities to a student network relying solely on radar data. The evaluation results demonstrate that the proposed CRKD-based model significantly improves the radar-only beam prediction performance. In particular, the student network substantially reduces the number of parameters compared to the teacher network, highlighting the effectiveness of our resource-efficient design. In summary, we make the following key contributions:

\begin{itemize}
    \item \textit{Multimodal realistic simulation framework:}  
    We integrate CARLA (to generate diverse sensor data) and MATLAB (to simulate mmWave channel communication) to create a virtual environment that accurately reflects real-world conditions, allowing robust performance evaluations.

    \item \textit{Cross-modal relational knowledge distillation:}  
    \textcolor{black}{We introduce a new beam prediction–specialized method that transfers relational features from a multimodal teacher model (trained with LiDAR, radar, GPS, and RGB) to a student with only radar. This approach preserves predictive accuracy while reducing sensor dependencies and computational complexity.}

    \item \textit{Analysis of generated multimodal data:}  
    We analyze the generated dataset. This analysis reveals the specific characteristics of the dataset, such as the distribution of beam indices across the entire dataset and the increased complexity of multi-lane scenarios with multiple strong signal paths. These insights highlight key challenges in beam prediction, such as dealing with skewed label distributions and maintaining accuracy in environments with higher multi-path variability.

    \item \textit{Extensive performance evaluation:}     
    We validate our approach using top-$k$ accuracy, mean received signal strength (RSS) and mean percentile rank (MPR) in urban scenarios with multi-lane. Results show that our radar-only student model achieves over $94\%$ of the teacher model's accuracy with only $10\%$ of the teacher network’s parameters. This demonstrates that cross-modal distillation can effectively preserve predictive performance under strict resource constraints, even in complex environments with high multipath and mobility.
\end{itemize}

The rest of this paper is organized as follows. Section~\ref{system_model} introduces the proposed multimodal sensing-based beam prediction system for multiple vehicles. Section~\ref{proposed_framwork} presents the realistic multimodal simulation framework, which is designed to generate multimodal sensing training data. In Section~\ref{main_method}, we propose CRKD for training a single-modal beam prediction model. Section~\ref{exp_results} analyzes the simulation results and, finally, Section~\ref{conclusion} concludes the paper with key findings and future directions.

\begin{figure*}[t!]
    \centering
    \includegraphics[width=0.9\textwidth]{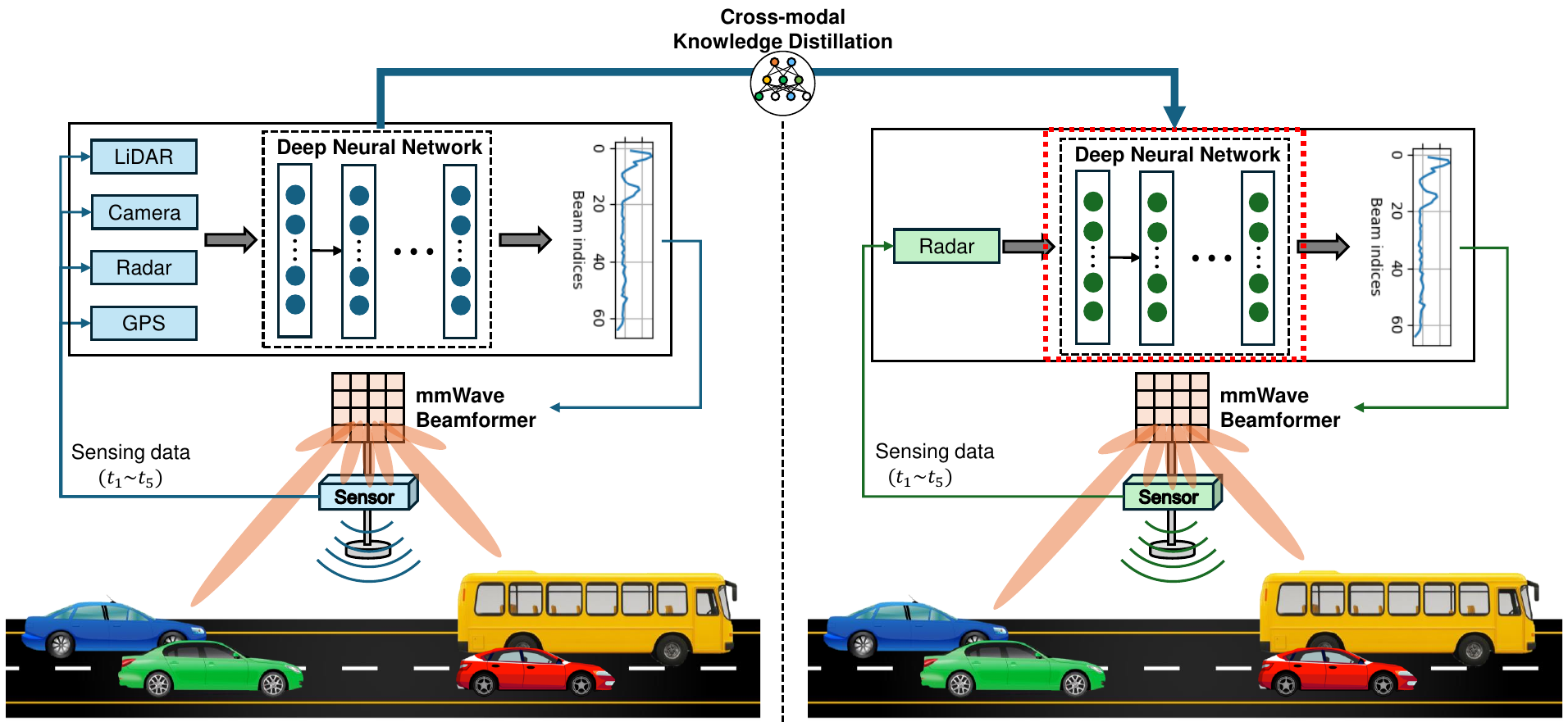}
    \caption{System model for multi-beam prediction with cross-modal knowledge distillation.}
    \label{simulation_active}
\vspace{-.175in}
\end{figure*}
\section{System model}
\label{system_model}
As illustrated in Fig.~\ref{simulation_active}, we consider multimodal sensing-based beam prediction for downlink mmWave communications in multiple V2I environments, which consists of a mmWave beamformer with a uniform rectangular array (URA) system of \(N\) antennas and a set \(\mathcal{V}\) of \(V\) vehicles, each with a single antenna. The beamformer employs a predefined beam codebook \(\mathcal{B} = \{ c_{1},\cdots,c_{B} \}\) of size \(B\), where each \(c_b\) corresponds to a beam pattern realized by a weight vector \(\boldsymbol{w}_b \in \mathbb{C}^{N}\). The beamformer selects the next optimal future beam index using a deep neural network (DNN) based on the sensing data from the previous observation window of size \(P\). The sensing information at the channel sampling interval \(t \in \mathcal{T}\) can be written as \(\mathcal{X} = \{ x[t-P+1],\cdots,x[t] \}\). We also consider two types of beamformers with different sensing configurations:
\begin{itemize}
    \item \emph{Multimodal Beamformer} (equipped with LiDAR, radar, GPS, and camera):
    This beamformer can extract rich environmental features such as 3D object shapes, distances, velocity information, and approximate location coordinates. This multimodal approach can produce highly accurate beam predictions, but requires significant hardware and computational resources.
    \item \emph{Radar-Only Beamformer}:  
    This is a beamformer that is restricted to radar measurements, which are generally lower-cost and robust under various environmental conditions (e.g., fog, rain). However, it may lack certain positional or visual details that LiDAR or cameras could provide, potentially reducing its beam prediction accuracy if used in isolation.
\end{itemize}
Despite these differences in sensing complexity, our goal is to develop a learning-based beam prediction method that minimizes performance loss when only radar sensing is available.

\color{black}
\subsection{Network model}
To model the characteristics of mmWave communication, we adopt a geometric ray-tracing-based channel framework that captures sparse multipath, directional propagation, and frequency-selective path loss. All path-level parameters $\{\beta_{v,l}, \theta_l, \phi_l, \tau_l\}_{l=1}^L$ are generated using MATLAB’s \texttt{raytrace} function, which implements a 3D shoot–bounce–ray (SBR) solver consistent with 3GPP TR 38.901. This setup includes specular reflections, single-edge diffractions (up to order 2), and realistic urban materials (e.g., concrete, glass, metal), modeled in an urban microcell environment imported from CARLA.

When applying a beamforming weight vector $\boldsymbol{w}_b$ of a beam pattern $c_b \in \mathcal{B}$, the magnitude of the response of the array for a vehicle $v$ over the path $l$ is calculated as:

\begin{equation}
R^{\mathrm{tx}}_{v,l}(c_b) = 10\log_{10} \left| \boldsymbol{w}_{b}^\mathrm{H} \cdot \boldsymbol{a}_{v}(\theta_{l},\phi_{l}) \right|,
\label{equ:responce}
\end{equation}

where $\boldsymbol{a}_{v}(\theta_{l},\phi_{l}) \in \mathbb{C}^N$ is the array response vector at azimuth $\theta_{l}$ and elevation $\phi_{l}$, and $^\mathrm{H}$ denotes the Hermitian (conjugate transpose). This response reflects the directional gain of the array for a specific path.

The total received signal strength (RSS) for vehicle \(v\) using beam pattern \(c_b\) is determined by aggregating contributions from all \(L\) propagation paths, weighted by path losses:

\begin{equation}
S_{v}(c_b) 
= \sum_{l=1}^{L} \left( R^{\mathrm{tx}}_{v,l}(c_b) - P^{l}_{v} \right),
\end{equation}

where \(P^{l}_{v}\) represents the path loss of the $l$-th path. This loss includes distance-based attenuation, material-dependent reflection loss, and large-scale fading. A common empirical model expresses this as follows:

\begin{equation}
P_{v}^l (\mathrm{dB}) 
= P_0 + 10\alpha \log_{10}\!\left( d_{v,l} \right) + \chi_{\sigma},
\end{equation}

where \(P_0\) is the reference path loss at 1 meter, \(\alpha\) is the path loss exponent, \(d_{v,l}\) is the distance of the $l$-th path, and \(\chi_{\sigma}\sim\mathcal{N}(0,\sigma^2)\) models shadow effects.

In addition, the effective channel gain for each path includes a carrier-dependent phase shift due to propagation delay $\tau_{v,l}$, expressed as:

\begin{equation}
h_{v,l} = \sqrt{\beta_{v,l}}\,e^{-j2\pi f_c \tau_{v,l}},
\label{eq:channel_gain}
\end{equation}

which is embedded in $\beta_{v,l}$ during the simulation. This realistic channel model reflects the sparse, high-directivity nature of mmWave environments and forms the foundation for evaluating beam prediction accuracy under our proposed knowledge distillation framework. Importantly, this formulation ensures that beam prediction models are trained and evaluated under physically realistic and path-dependent conditions, allowing them to learn angular selectivity and NLoS behavior intrinsic to real-world mmWave propagation.

\color{black}
\subsection{Problem statement}
We will investigate how to effectively leverage the sensing information obtained from the beamformer sensors for the mmWave beam prediction problem. We aim to select an optimal beam pattern \(c_b\) from the candidate beams in the codebook $\mathcal{B}$ that maximizes the sum of RSS of vehicles. In a practical mmWave downlink scenario, the base station (or beamformer) observes the dynamic environment and selects the beam pattern \(c_b\) to maximize received strength across vehicles. For the effective channel \(\boldsymbol{h}_{v}\) of vehicle \(v\), the weight vector \(\boldsymbol{w}_b\in\mathbb{C}^{N}\) in \(\mathcal{B}\) should be chosen to achieve a high inner product \(\boldsymbol{h}_{v}^H \boldsymbol{w}_b\). The challenge is that \(\boldsymbol{h}_{v}\) evolves quickly due to mobility, blockage, and reflections at mmWave frequencies, leading to frequent re-selection of the beam pattern. Exhaustive beam sweeping can be performed to identify \(c_b^{*}\), but this approach is time-consuming, particularly for large \(|\mathcal{B}|\). Therefore, sensing-aided beam prediction that uses radar, LiDAR, GPS, or camera data can help to infer the optimal beam directly from environmental observations. However, to train learning models for beam prediction, we require comprehensive multimodal data, which we address in Section~\ref{proposed_framwork}.

Hence, our goal is to develop a deep learning framework capable of predicting beams using the collected sensory data $\mathcal{X}$. We consider two different models: one that uses multimodal sensing data and another that relies on radar only. The intended output of these deep learning models is a probability distribution $\boldsymbol{p} = [p_1, p_2, \ldots, p_B]$ over the beamforming codebook $\mathcal{B}$. The beam pattern predicted by the model with the highest prediction probability is given by:

\begin{equation}
\hat{\boldsymbol{b}} = \arg\max^{\mathcal{B}}_{b}p_{b}.
\end{equation}
For multi-vehicle scenarios, one may sum or average \(\mathrm{RSS}_{v}(c_b)\) across all vehicles \(v \in \{1, \dots, V\}\) to evaluate the utility of each beam pattern for the entire coverage area. Identifying the index corresponding to the highest sum of RSS across multiple vehicles, the optimal beam pattern is given by:

\begin{equation}
\boldsymbol{b}^{*} = \arg\max^{\mathcal{B}}_{c_{b}}\sum^{\mathcal{V}}_{v=1}S_{v}(c_b).
\end{equation}
which gives us the beam pattern \(c_b\) that maximizes the cumulative signal strength across all vehicles.

The beam prediction model $f(\cdot)$ is parameterized by a set of parameters $\Theta$. These parameters are learned from the training dataset $\{ \mathcal{X}, \boldsymbol{b}^{*} \}$ at the channel sampling interval $t$, which contains the sensing information along with the corresponding optimal beam patterns. Consequently, the optimization problem predicting the beam for the future channel sampling interval $t+1$ can be written as \cite{jiang2022lidar}:

\begin{equation}
f^{*}(\mathcal{X}, t+1; \Theta^{*}) = \underset{f(\cdot),\Theta}{\arg\max~} \mathbb{P} \left\{ f(\mathcal{X}, t+1; \Theta) = \boldsymbol{b}^{*}[t+1] \right\}.
\end{equation}

The prediction models are referred to as $f_{\textrm{multi}}(\cdot)$ with parameter $\Theta_{\textrm{multi}}$, which uses multimodal sensing information, and $f_{\textrm{mono}}(\cdot)$, which uses radar-only sensing information, depending on the type of sensor used. Next, we present how to generate training data and learn multimodal models.

\section{Multimodal Realistic Simulation Framework for Sensing-aided Communication}
\label{proposed_framwork}
\begin{figure}[t]
    \centering
    \includegraphics[width=\columnwidth]{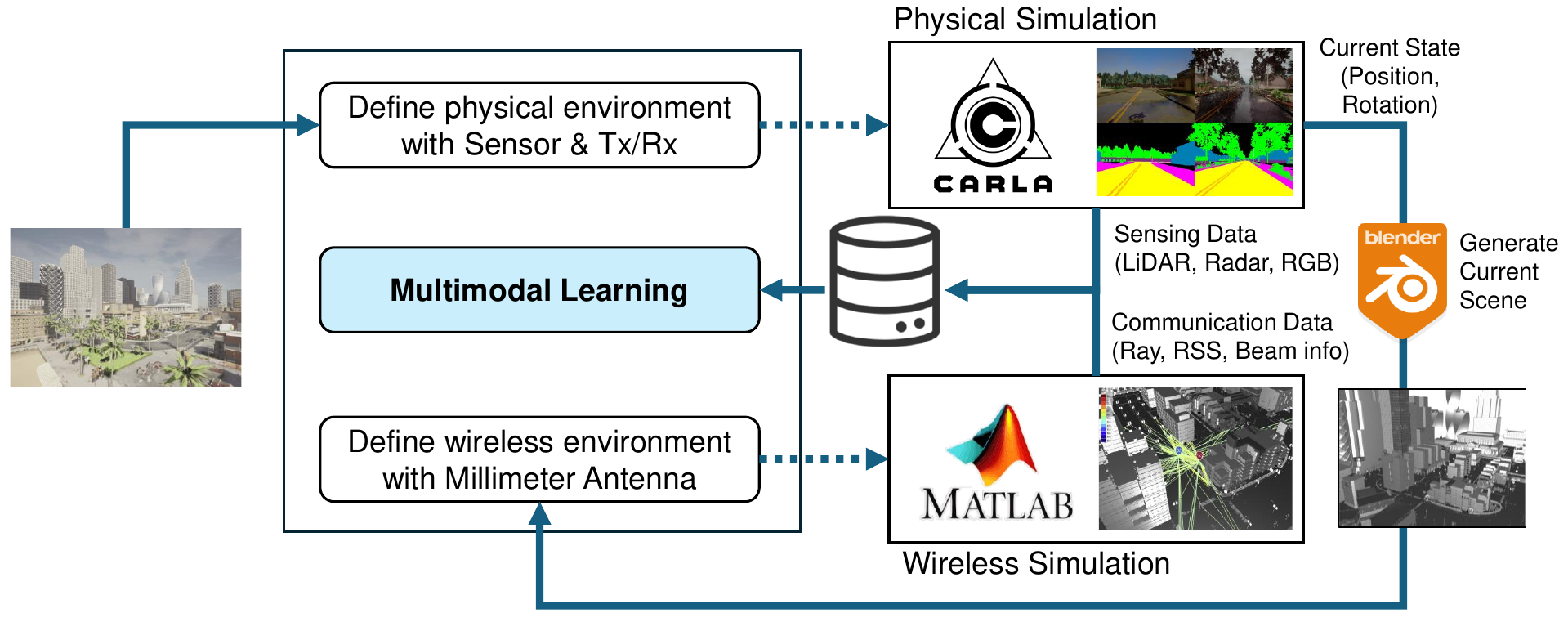}
    \caption{Multimodal realistic simulation framework based on autonomous driving tool CARLA and MATLAB.}
    \label{fig:framework}
\vspace{-.175in}
\end{figure}
We now present a realistic multimodal simulation framework designed to generate multimodal sensor data and perform wireless communication experiments in a detailed virtual urban environment. Fig.~\ref{fig:framework} illustrates the workflow of the proposed framework. We use the autonomous driving tool CARLA \cite{Dosovitskiy17} to generate realistic sensing information, while MATLAB \cite{MATLAB} is used for communication experiments. Next, we describe the overall workflow, including procedures for sensor data generation and digital environment reconstruction. We then discuss how the reconstructed environment supports multimodal sensor simulation and ray-tracing-based channel generation, ultimately producing the data required for cross-modal knowledge distillation.

The proposed framework consists of four main stages: environment setup, sensing data generation, 3D map reconstruction, and wireless channel simulation. By integrating CARLA’s robust vehicular and sensor modeling capabilities with MATLAB’s communication toolboxes, we ensure realistic modeling of both sensor signals and mmWave propagation characteristics.

\textbf{Environment Setup:} To simulate realistic urban scenarios, we place multiple vehicles and one or more base stations within the CARLA environment. The built-in traffic control and navigation functions of CARLA govern the autonomous movements of these vehicles, adhering to traffic signals, speed limits, and road geometry. This configuration provides dynamic mobility patterns and a diverse range of sensing conditions to evaluate beamforming and channel characteristics:
\begin{itemize}
    \item \emph{Base Station Placement:} We position the base station(s) at fixed points, such as roadside units or rooftop installations, consistent with typical urban deployments.
    \item \emph{Vehicle Distribution:} Vehicles are spawned in random locations or traffic centers, allowing a variety of relative positions and velocities for a more comprehensive data collection.
    \item \emph{Environmental Dynamics: Here,} changes in lighting, weather, and traffic density have been introduced to simulate different conditions (e.g., night, fog, heavy traffic).
\end{itemize}

\textbf{Sensing Data Generation:} To capture the information required for subsequent communication analysis, we instantiate multiple sensors at the base station, as well as in vehicles if needed. Using CARLA’s APIs, we configure sensor modalities such as:
\begin{itemize}
    \item \emph{LiDAR:} \textcolor{black}{It} generates 3D point clouds, providing high-resolution distance and object shape information.
    \item \emph{Radar:} \textcolor{black}{It} offers lower-resolution distance and velocity measurements, robust in adverse weather or lighting conditions.
    \item \emph{RGB Cameras:} \textcolor{black}{It provides} rich color image data for visual context (e.g. obstacle detection, object classification).
    \item \emph{GPS:} \textcolor{black}{It} logs positional coordinates and velocities.
\end{itemize}
This multimodal data captures the dynamic movement of vehicles and environmental details such as buildings, roads, and other objects. \textcolor{black}{Then, all sensing} data are synchronized in time, ensuring consistent data alignment across LiDAR, radar, and camera outputs.

\textbf{Digital 3D Reconstruction:} Although CARLA renders a realistic environment for autonomous driving simulations, \textcolor{black}{it is required} MATLAB for the compatible 3D model format to conduct wireless channel simulations. To harmonize these platforms, we perform a conversion of the CARLA maps using Blender API:
\begin{itemize}
    \item \emph{Map Export and Conversion:} We export the CARLA environment, including roads and buildings, into an intermediate 3D file format (e.g., \texttt{FBX} or \texttt{OBJ}). 
    \item \emph{Scripting in Blender:} The Blender API is used \textcolor{black}{for scripting} the conversion process, ensuring that the geometry, coordinates of the texture, and scale of the model are preserved.
    \item \emph{MATLAB Import:} The resulting 3D file is then imported into MATLAB, generating a mesh-based environment consistent with the virtual scene in CARLA.
\end{itemize}
This process guarantees that the geometry, dimensions, and layout remain accurate on both simulation platforms.

\label{proposed_method}
\begin{figure*}[t]
    \centering
    \includegraphics[width=0.9\textwidth]{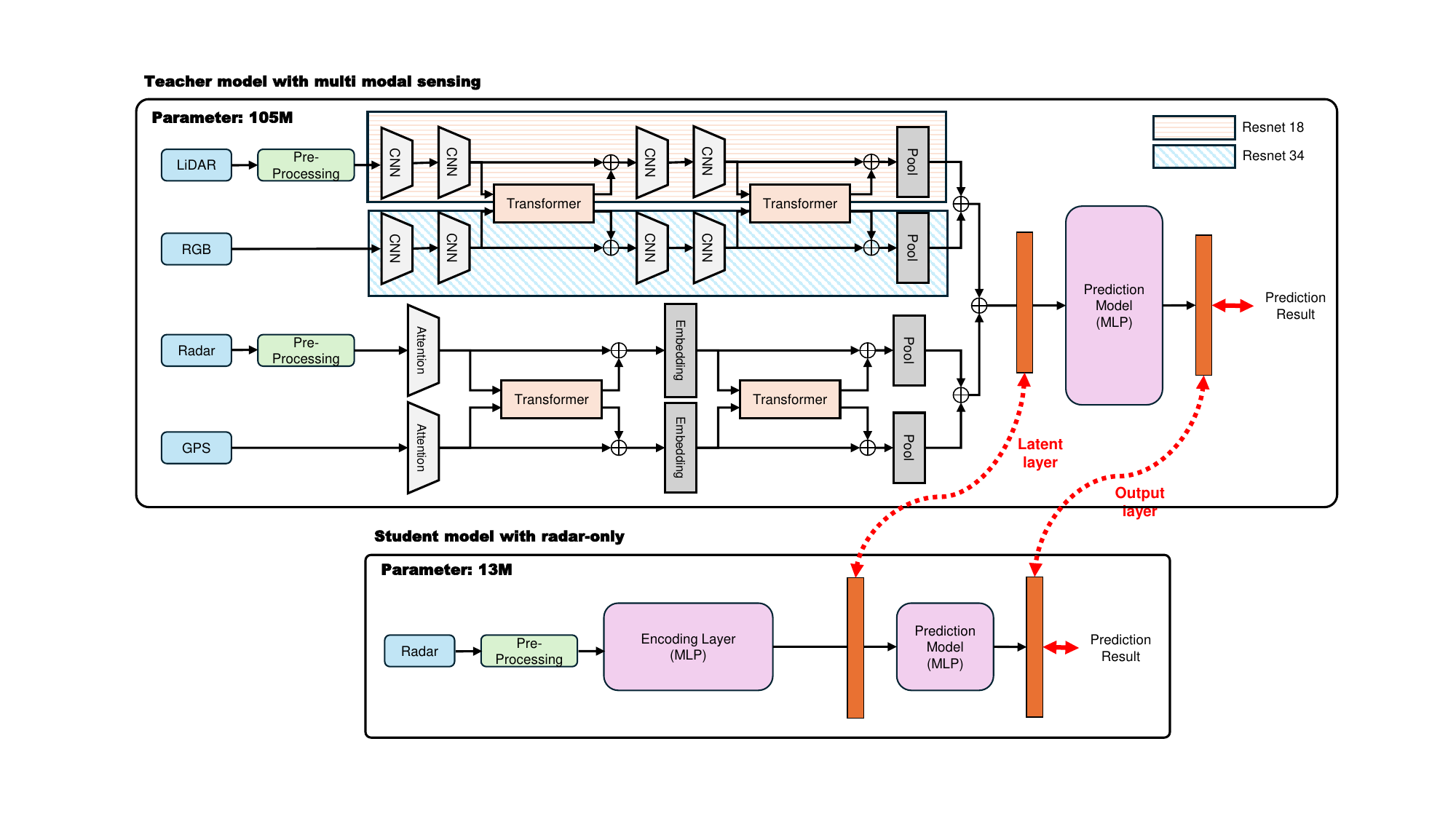}
    \caption{The proposed structure of cross-modal knowledge distillation from multimodal (LiDAR, RGB, radar, and GPS) to monomodal (radar).}
    \label{fig:kd}
\vspace{-.175in}
\end{figure*}

\textbf{Wireless Channel Simulation:} After importing the 3D environment into MATLAB, we perform ray-tracing-based wireless channel simulations to capture the propagation characteristics of mmWave. By tracing signal paths, reflections, diffractions, and line-of-sight (LoS) or non-line-of-sight (NLoS) components, we gain detailed insights into how each beam pattern interacts with the reconstructed urban scene. Specifically:
\begin{itemize}
    \item \emph{Ray-Tracing Algorithm:} Multiple rays are cast from the base station in different directions, and their interactions with objects are calculated based on reflection, scattering, or diffraction coefficients.
    \item \emph{Beam Pattern Evaluation:} \textcolor{black}{ The received signal strength (RSS) is evaluated for each codebook beam at various vehicle positions, creating a labeled dataset linking spatial and sensing data to channel observations.}
    \item \emph{Dynamic Updates:} \textcolor{black}{As vehicles move, updated positional data from CARLA can be used to re-simulate or predict channel states, thus creating a time-series dataset of multimodal sensing and wireless measurements.}
\end{itemize}

We \textcolor{black}{integrate} the multimodal sensor data obtained from CARLA with the corresponding RSS and channel parameters from the MATLAB ray-tracing simulations. This fused dataset includes:
\begin{itemize}
    \item \emph{Sensor Streams (LiDAR, radar, RGB, GPS):} High-dimensional observations of the environment.
    \item \emph{Channel State Information (Ray-Tracing):} Path loss, delay spread, angle of arrival (AoA), angle of departure (AoD), and RSS for each beam pattern.
    \item \emph{Temporal and Positional Labels:} Timestamps, vehicle IDs and spatial coordinates to enable sequential or spatial modeling.
\end{itemize}
These data \textcolor{black}{are} the foundation for training and validating beam prediction algorithms, including the proposed cross-modal knowledge distillation approach. 

In summary, the proposed multimodal simulation framework (Fig.~\ref{fig:framework}) seamlessly integrates autonomous driving and wireless communication simulations by generating dynamic and realistic urban scenarios through CARLA’s traffic control and sensor suite, reconstructing the resulting 3D environment in MATLAB for accurate ray-tracing analysis, and creation of a comprehensive dataset that encompasses both sensor observations and channel measurements. This unified approach lays the foundation for rigorous performance evaluation of sensing-aided beam prediction and enables advanced methods such as cross-modal knowledge distillation, which leverages rich multimodal data during training while reducing sensor dependencies at inference.

\section{Cross-modal Relational Knowledge Distillation From multimodal to monomodal}
\label{main_method}
As illustrated in Fig.~\ref{fig:kd}, our objective is to distill the knowledge from a teacher network trained on multimodal sensing information (LiDAR, radar, GPS, and RGB) into a student network that relies only on radar data. Our teacher model uses a transformer-based sensor fusion (with a ResNet backbone for images), resulting in $105$ million parameters. \textcolor{black}{In contrast,} the student \textcolor{black}{network} is a radar-only multilayer perceptron (MLP) with $13$ million parameters. In this section, we outline the preprocessing steps for the multimodal data and the training process of the teacher network. We then describe how the student network is trained using our cross-modal knowledge distillation.

\vspace{-.175in}
\subsection{Multimodal preprocessing}
The teacher network, captured by \(f_{\mathrm{multi}}(\cdot)\), has four sensor modalities: LiDAR, radar, GPS, and RGB images. Each sensor type is pre-processed in a format that can be consistently fed into subsequent encoders:

\begin{itemize}
    \item \emph{LiDAR (Bird’s‐Eye View \textcolor{black}{(BEV)}):} Raw LiDAR point clouds \(\bigl[\text{x},\text{y},\text{z},\text{intensity}\bigr]\) are projected onto a 2D plane to form a BEV representation. This transformation highlights object occupancy and relative positioning in a top‐down image layout, filtering out ground reflections, and simplifying 3D geometry into 2D grids.
    \item \emph{Radar (Highest Point Sampling (HPS)):} To handle radar data \(\bigl[\text{velocity},\,\text{azimuth},\,\text{altitude},\,\text{depth}\bigr]\), we apply HPS, which downsamples radar point clouds to a fixed size, while ensuring consistency across different sampling intervals or noise conditions.
    \item \emph{GPS:} GPS coordinates are recorded as numerical features. These can be fed into fully connected layers or concatenated with other high‐level features to enrich the spatial context.
    \item \emph{RGB  \textcolor{black}{Images}:} Camera images are processed with a standard 2D convolutional neural network (CNN), such as a ResNet block, producing visual feature maps that capture color and texture information.
\end{itemize}

\subsection{Learning of the teacher network}
\subsubsection{Transformer-Based Fusion:}
\textcolor{black}{After separate backbone encoders extract the LiDAR/RGB and radar/GPS features, the resulting embeddings are fused using a \emph{Transformer} module that captures cross‐modal correlations:}
\begin{itemize}
    \item \emph{Image‐Based Encoding:} LiDAR‐BEV and RGB image features each pass through their own CNN backbone (e.g., ResNet). The output feature maps are flattened or pooled and then fed into a Transformer that learns global correlations among spatial patches.
    \item \emph{Point‐Based Encoding:} Radar and GPS data undergo an attention-based layer to compress point‐wise features, followed by an attention layer and a linear layer to reduce dimensionality. These intermediate embeddings are then processed by the same (or parallel) Transformer to align with the image‐based features.
\end{itemize}
By applying multi-head attention to the connected tokens of LiDAR, radar, RGB, and GPS, the Transformer separates and merges point-based and image-based data, resulting in a unified multimodal representation \(\boldsymbol{Z}_{\mathrm{multi}}\) that provides greater awareness of the environment.

\subsubsection{Beam Prediction and Focal Loss:}
The fused representation \(\boldsymbol{Z}_{\mathrm{multi}}\) is finally passed to a prediction layer to output a probability distribution on the beamforming codebook \(\mathcal{B}\). The prediction model $f_{\mathrm{multi}}(\cdot)$ uses a focal loss function and uses stochastic gradient descent (SGD) for optimization. Focal loss is a modification of the standard cross-entropy loss designed to address the class imbalance problem. In datasets with imbalanced classes, the majority class can dominate the loss, leading to poor performance for the minority class. The focal loss function is formulated as follows:

\begin{equation}
\label{eq:focal_loss}
L_{\mathrm{focal}}
= 
-\bigl(1 - p_{\boldsymbol{b}^{*}}\bigr)^{\gamma}
\log\bigl(p_{\boldsymbol{b}^{*}}\bigr),
\end{equation}
where \(p_{b}\) is the predicted probability of selecting beam \(c_{b}\). \(\boldsymbol{b}^{*}\) is the ground‐truth beam index (i.e., the beam maximizing the sum‐RSS), and \(\gamma = 2\) is a focusing parameter in our experiments. In highly imbalanced datasets where certain beams dominate, the focal loss ensures more attention is given to challenging samples. The teacher network uses only the focal loss in (\ref{eq:focal_loss}) as a loss function.

\subsection{Learning of the student network}
The student network, \(f_{\mathrm{mono}}(\cdot)\), operates exclusively on radar input. The student network is a feedforward multilayer perceptron (MLP) with roughly 13 million parameters by default. The MLP of the student consists of $6$ fully connected layers with ReLU activations. This monomodal design reduces sensor and computational overhead, but naturally yields less environmental awareness than the multimodal teacher. Training \(f_{\mathrm{mono}}\) purely with a label-based loss often leads to suboptimal beam predictions. Hence, we apply KD framework to significantly enhance the student's performance while reducing the overall model size. 

\subsubsection{Conventional Knowledge Distillation:}
Conventional KD aims to transfer knowledge from the teacher to the student network by minimizing the loss of distillation. The distillation loss is calculated based on the difference between the features of the teacher and student networks. To calculate this difference, the conventional KD methods use the Kullback-Leibler (KL) divergence, which is a statistical measure that quantifies how a feature distribution $\mathcal{F}_{\mathrm{mono}}$ of the student network differs from a feature distribution $\mathcal{F}_{\mathrm{multi}}$ of the teacher network. For two feature distributions $P$ and $Q$, the KL divergence is calculated as \cite{kullback1951information}:

\begin{equation}
L_{\mathrm{kl}}(P||Q) = -T^{2}\sum^{F}_{f=1}\sigma(P)\log \left(\frac{\sigma(P)}{\sigma(Q)}\right),
\label{kldivergence}
\end{equation}
where $T$ is the temperature to control the distribution over features, to essentially smooth the distribution, thereby capturing the nuanced relationships between different features as learned by the teacher \textcolor{black}{network}. In our experiments, we set $T=2$. $F$ is the
total number of features. $\sigma(z)={e^{{z}_i}}/{\sum_j e^{{z}_j}}$ is the softmax function, where ${z}_i$ is the $i$-th element of the input vector $z$. 

To obtain more fine-grained knowledge from the teacher network, we incorporate not only the loss based on the final output features $\mathcal{F}^{\mathrm{end}}$ but also an additional loss term derived from the difference between the latent features $\mathcal{F}^{\mathrm{mid}}$ in the encoding layers. As a result, the overall loss for the student network, including both the original label loss and the distillation loss, can be calculated as:

\begin{equation}
\mathcal{L}_{\mathrm{kd}} = (1-\alpha)L_{\mathrm{focal}}\\ + \alpha
\sum_{l\in\{\mathrm{mid},\,\mathrm{end}\}}
L_{\mathrm{kl}}\!
\Bigl(
f_{t}^{(l)},\,f_{s}^{(l)}
\Bigr),
\label{eq:loss_st}
\end{equation}
where $\alpha$ is a weight parameter that balances the importance of the original loss and the distillation loss. \(f_{t}^{(l)}\) and \(f_{s}^{(l)}\) are the teacher and student feature maps at layer \(l\).

\begin{figure*}[!t]
        \centering
        \begin{subfigure}[t]{0.45\textwidth}
        \centering
               \captionsetup{justification=raggedright,singlelinecheck=false}
                \includegraphics[width=\linewidth]{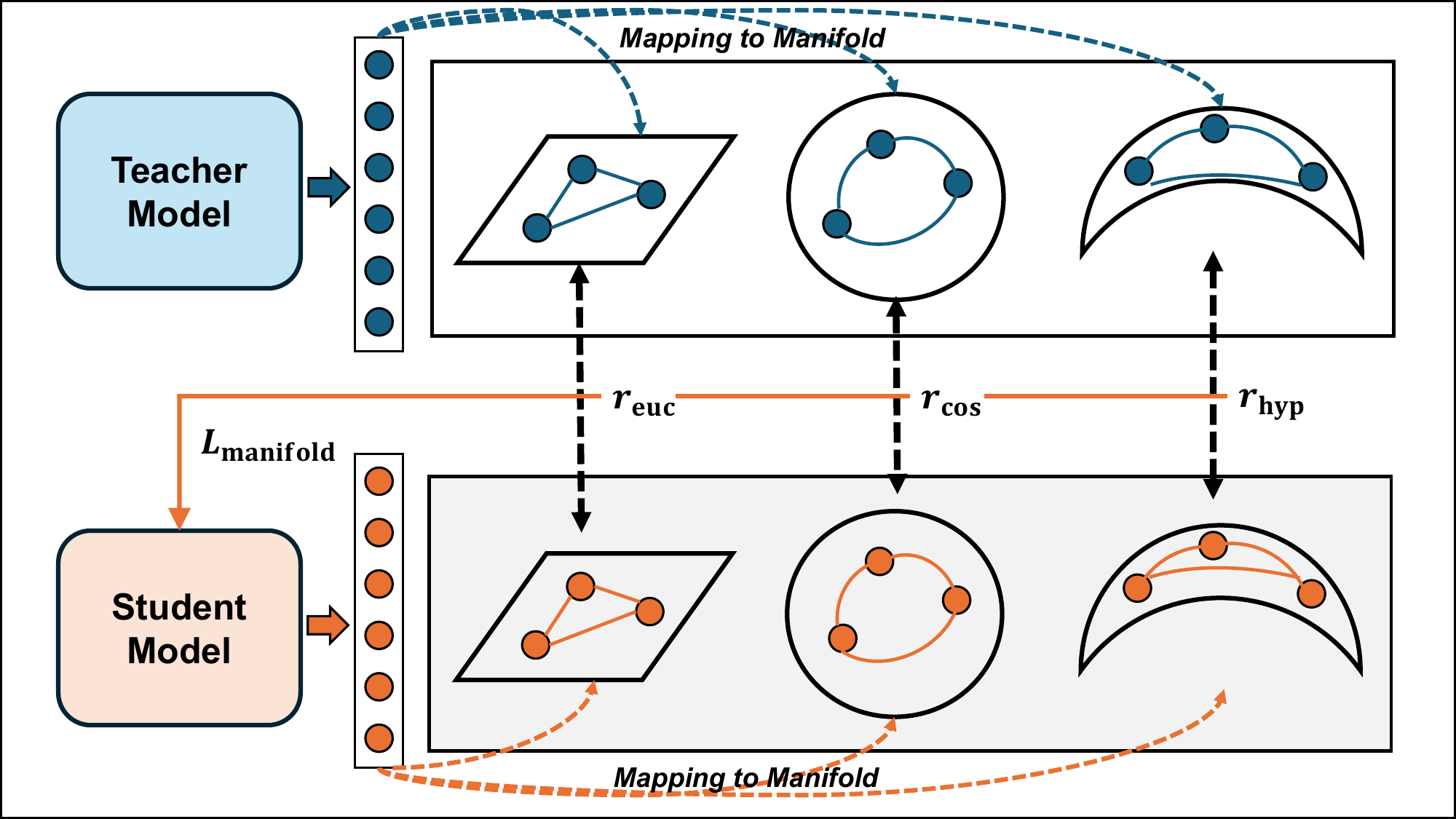}
                \caption{Manifold-Based Relationship.}
                \label{manifold}
        \end{subfigure}%
        \begin{subfigure}[t]{0.45\textwidth}
        \centering
               \captionsetup{justification=raggedright,singlelinecheck=false}
               \includegraphics[width=\linewidth]{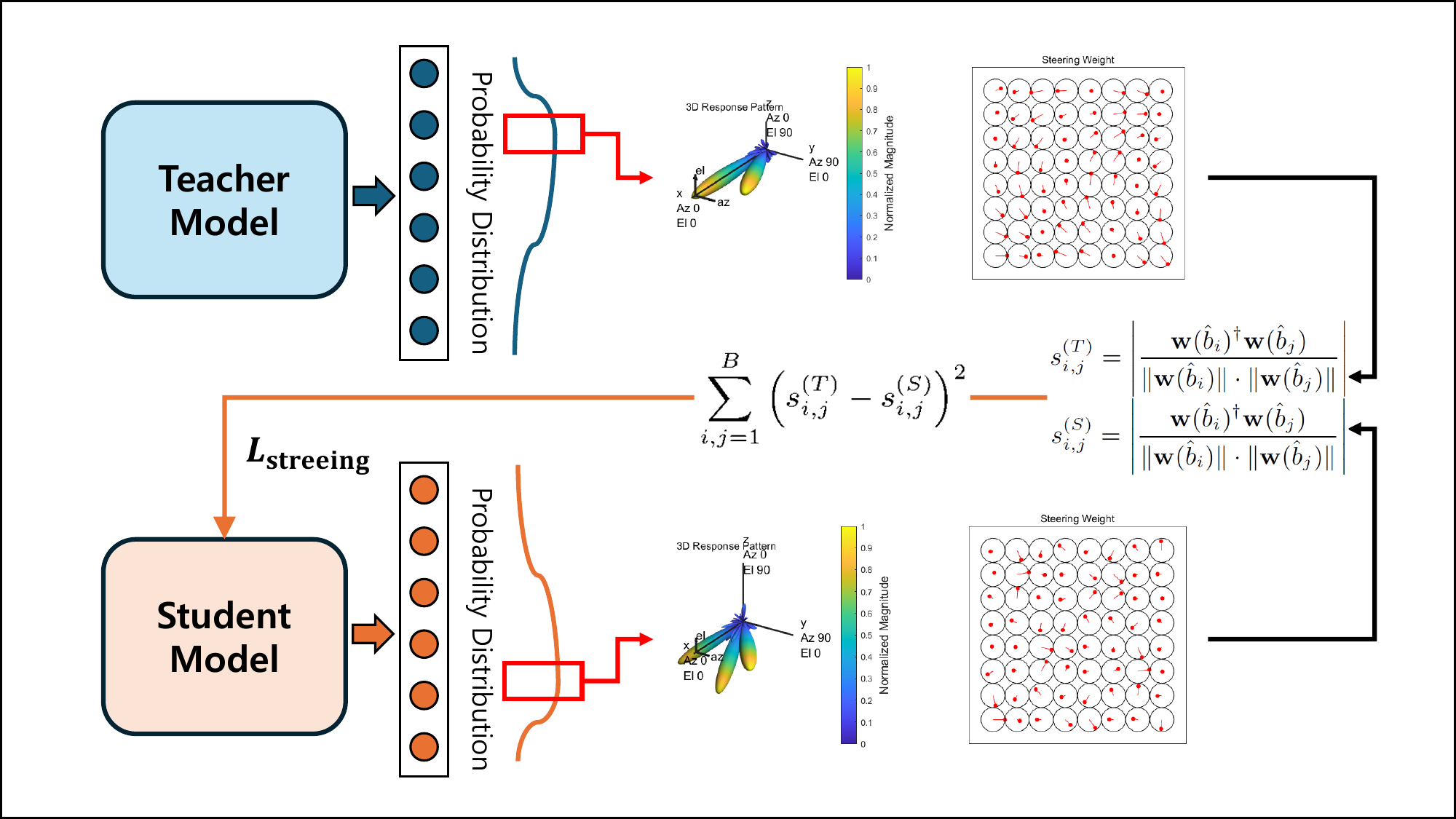}
                \caption{Beam Steering based Relationship.}
                \label{steeing}
        \end{subfigure}%
        \caption{Relational Knowledge Distillation with different relationships.}\label{fig:relational}
\vspace{-.175in}
\end{figure*}

\subsubsection{Relational Knowledge Distillation:}
\label{subsubsec:rkd}
\textcolor{black}{While conventional KD focuses on aligning outputs or feature representations on a per-sample basis, \emph{relational knowledge distillation} (RKD) goes a step further by preserving \emph{pairwise relationships} between samples in the feature space. This approach is especially beneficial in cross-modal knowledge distillation, where the teacher model operates on richer multimodal data while the student has access to only a single modality (i.e., radar). In our framework, we implement two methods of relational knowledge distillation: (a) a manifold-based loss that aligns the latent feature geometry between teacher and student, and (b) a novel beam steering-based loss that captures the relational structure in the complex-valued beamspace at the output level. These complementary losses enable the student to better mimic the teacher’s internal representations and output behaviors under sensor constraints.}

\paragraph{Manifold‐Based Relationship Measures:}
To extract deep features from various relationships of the latent layer, we adopted a method called DistilVPR \cite{wang2024distilvpr}. To capture higher-order relationships between feature embeddings, DistilVPR computes pairwise distances or similarities in three distinct manifolds: Euclidean (flat), spherical (positive curvature), and hyperbolic (negative curvature). For any two feature vectors $\boldsymbol{t}_i$ and $\boldsymbol{t}_j$, we denote:

\[
r_{\mathrm{euc}}(\boldsymbol{t}_i,\boldsymbol{t}_j),\quad
r_{\mathrm{cos}}(\boldsymbol{t}_i,\boldsymbol{t}_j),\quad
r_{\mathrm{hyp}}(\boldsymbol{t}_i,\boldsymbol{t}_j).
\]

The Euclidean distance (or $\ell_2$) is the most common metric to measure pairwise dissimilarity in a flat manifold as follows:

\begin{equation}
\label{eq:reuc}
r_{\mathrm{euc}}(\boldsymbol{t}_i,\boldsymbol{t}_j)
= \bigl\|\boldsymbol{t}_i - \boldsymbol{t}_j\bigr\|_{2}
= \sqrt{\sum_{d=1}^{D} \bigl(t_{i,d} - t_{j,d}\bigr)^2},
\end{equation}
where $\boldsymbol{t}_i,\boldsymbol{t}_j \in \mathbb{R}^D$ and $D$ is the feature dimension. In the context of knowledge distillation, matching these Euclidean distances in the teacher and student feature spaces ensures that if two samples $(i,j)$ are close in the teacher's embeddings, they remain close in the student's embeddings. While Euclidean distance captures raw feature differences, the cosine similarity is more sensitive to angular relationships and is often interpreted as measuring distance on a spherical manifold. To explore the spherical-based relationship, the cosine distance is given by:
\begin{equation}
\label{eq:rcos}
r_{\mathrm{cos}}(\boldsymbol{t}_i,\boldsymbol{t}_j)
= \frac{\langle \boldsymbol{t}_i,\boldsymbol{t}_j\rangle}{\|\boldsymbol{t}_i\|\;\|\boldsymbol{t}_j\|},
\end{equation}
where $\langle \cdot,\cdot\rangle$ represents the inner (dot) product. In practice, one may use $1 - r_{\mathrm{cos}}$ as a distance-like measure. Cosine-based relationships are useful when the magnitude of vectors is less important than their direction, a scenario common in classification or retrieval tasks. \textcolor{black}{We incorporate a hyperbolic measure using the Poincar\'e ball model to capture hierarchical or tree-like structures and negative curvature.} When $\boldsymbol{t}_i, \boldsymbol{t}_j \in \mathbb{R}^D$, we can project each vector onto the Poincare space $\mathcal{D}_c^D$ (with curvature parameter $c>0$) via an exponential mapping as follows:

\begin{equation}
\label{eq:exp_map}
\boldsymbol{t}_i^{(\mathrm{hyp})}
= \exp_{\boldsymbol{0}}^c(\boldsymbol{t}_i)
= \tanh\!\Bigl(\sqrt{c}\,\|\boldsymbol{t}_i\|\Bigr)
\,\frac{\boldsymbol{t}_i}{\sqrt{c}\,\|\boldsymbol{t}_i\|}.
\end{equation}

Given two hyperbolic embeddings $\boldsymbol{t}_i^{(\mathrm{hyp})}, \boldsymbol{t}_j^{(\mathrm{hyp})} \in \mathcal{D}_c^D$, their hyperbolic distance $r_{\mathrm{hyp}}(\cdot,\cdot)$ is computed as:

\begin{equation}
\label{eq:rhyp}
\begin{aligned}
r_{\mathrm{hyp}}\bigl(\boldsymbol{t}_i,\boldsymbol{t}_j\bigr)
&= d_{\mathrm{hyp}}\Bigl(
\boldsymbol{t}_i^{(\mathrm{hyp})},\,
\boldsymbol{t}_j^{(\mathrm{hyp})}
\Bigr) \\
&= \frac{2}{\sqrt{c}}\;
\mathrm{arctanh}\!\Bigl(
\sqrt{c}\,\Bigl\|\,
-\boldsymbol{t}_i^{(\mathrm{hyp})}
\oplus_c\,
\boldsymbol{t}_j^{(\mathrm{hyp})}
\Bigr\|
\Bigr),
\end{aligned}
\end{equation}
where $\oplus_c$ is the M\"obius addition in the Poincar\'e ball. The negative curvature of hyperbolic space often helps preserve hierarchical relationships in feature embeddings. We then compare these teacher relations to the corresponding student relations, e.g., \(r_{\mathrm{euc}}(s_{i}, s_{j})\). One can define a relational loss over all pairs \((i,j)\) as:

\begin{align}
L_{\text{latent-rel}}
=
\sum_{i,j=1}^{B}
\Bigl[
&\,d\bigl(r_{\mathrm{euc}}(t_{i}, t_{j}),\,r_{\mathrm{euc}}(s_{i}, s_{j})\bigr)
\nonumber \\
&\quad
+\,d\bigl(r_{\mathrm{cos}}(t_{i}, t_{j}),\,r_{\mathrm{cos}}(s_{i}, s_{j})\bigr)
\nonumber \\
&\quad
+\,d\bigl(r_{\mathrm{hyp}}(t_{i}, t_{j}),\,r_{\mathrm{hyp}}(s_{i}, s_{j})\bigr)
\Bigr],
\label{eq:loss_rel}
\end{align}
where \(d(\cdot,\cdot)\) is a distance or divergence. This ensures that the student maintains the geometry of the higher order of the teacher, even if the absolute feature values differ because the teacher sees more modalities.

\color{black}
\paragraph{Beam Steering based Relationship Measures:}
In addition to the latent space structure, we propose a novel relational distillation loss in the output domain by leveraging the domain-specific properties of mmWave beamforming. Each predicted beam index corresponds to a complex-valued steering vector, which defines the phase alignment of signals transmitted across antenna elements. These vectors capture the directionality of beamforming and inherently encode spatial relationships between users or signal paths. Since steering vectors represent spatial directionality, aligning their pairwise similarities enables the student to learn beam-space relationships that are critical for mmWave propagation, where multipath components and direction selectivity play a dominant role in determining communication performance.

To preserve the relational structure of the beamspace, we compute a similarity matrix using the Hermitian inner product between normalized steering vectors. For two predicted beam indices $\hat{b}_i$ and $\hat{b}_j$ from the teacher model, their similarity is defined as:

\begin{equation}
s_{i,j}^{(T)} =
\left|
\frac{
\mathbf{w}(\hat{b}_i)^\dagger \mathbf{w}(\hat{b}_j)
}{
\|\mathbf{w}(\hat{b}_i)\| \cdot \|\mathbf{w}(\hat{b}_j)\|
}
\right|,
\label{eq:beam_sim}
\end{equation}

where $\mathbf{w}(\hat{b}_i) \in \mathbb{C}^N$ is the steering vector for the beam index $\hat{b}_i$, $^\dagger$ denotes the Hermitian (conjugate transpose), and $N$ is the number of antenna elements. The same operation is performed for the predicted beams of the student to yield $s_{i,j}^{(S)}$.

We then define a beamspace relational loss to align the structural similarities of beam directions between teacher and student:

\begin{equation}
L_{\text{output-rel}} =
\sum_{i,j=1}^{B}
\left(
s_{i,j}^{(T)} - s_{i,j}^{(S)}
\right)^2,
\label{eq:loss_beam_rel}
\end{equation}
where $B$ is the batch size. This loss ensures that, even without direct access to multimodal inputs, the radar-only student can mimic the teacher's understanding of directional similarity across samples. In scenarios with domain shifts or multiple strong paths, preserving this angular relational structure is critical for robust beam prediction. To the best of our knowledge, this is the first application of relational knowledge distillation based on complex-valued beam steering vector similarities for mmWave communication. While prior work like DistilVPR focuses on latent space geometry, our proposed loss explicitly leverages directional structure at the output level, a domain-specific innovation tailored to beamforming systems under sensor constraints. Consequently, we can replace the conventional KD term in \eqref{eq:loss_st} with relational loss as:

\begin{align}
\label{eq:cmkd_loss}
\mathcal{L}_{\mathrm{rkd}}
=
(1-\alpha)\,L_{\mathrm{focal}}
+
\alpha (L_{\text{latent-rel}} + L_{\text{output-rel}}),
\end{align}
Similar to the conventional KD loss function (\ref{eq:loss_st}), the parameter $\alpha$ balances the importance between the original supervised loss and the relational distillation losses.

\begin{figure*}[!t]
        \centering
        \begin{subfigure}[t]{0.45\textwidth}
        \centering
               \captionsetup{justification=raggedright,singlelinecheck=false}
                \includegraphics[width=\linewidth]{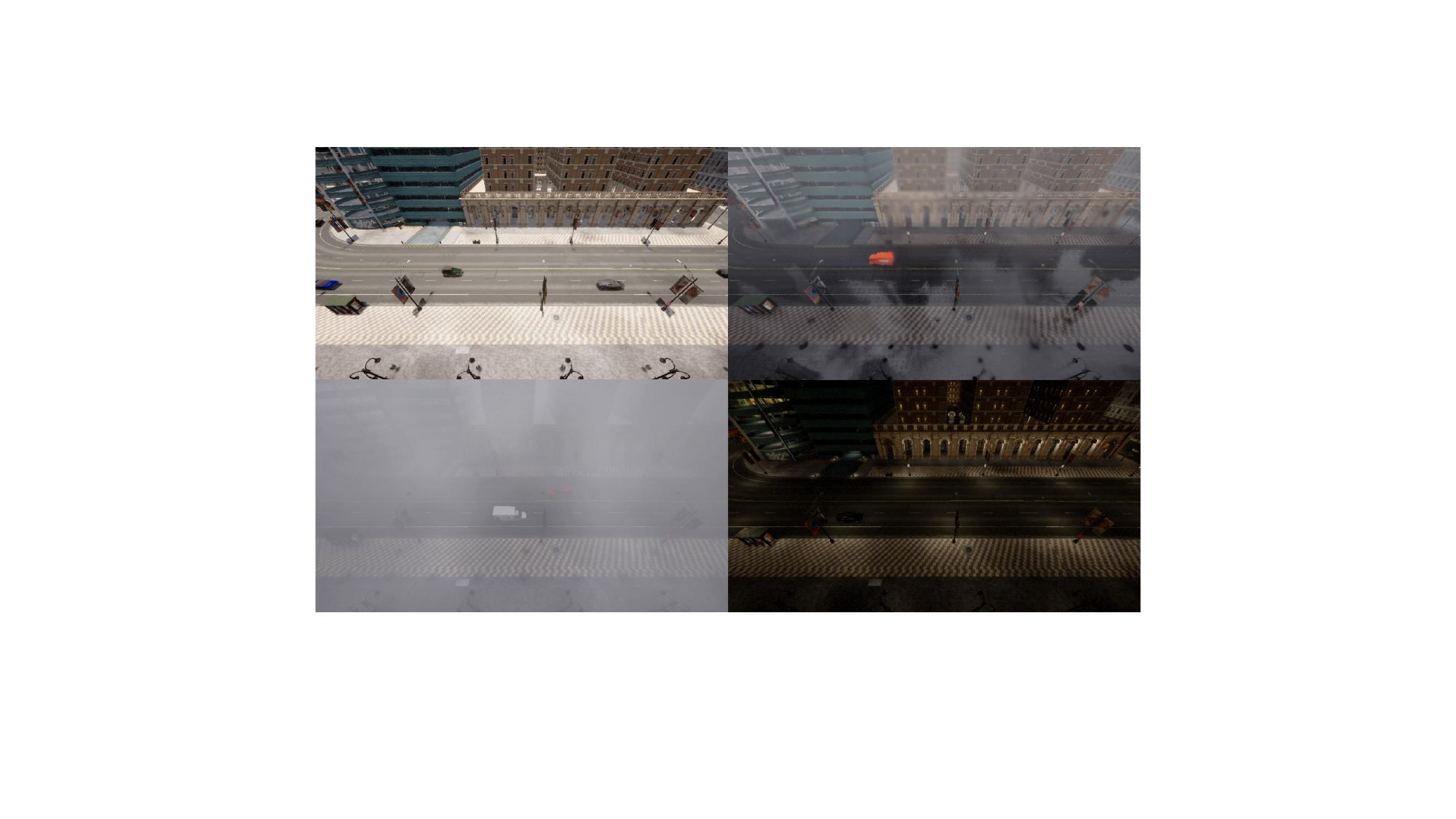}
                \caption{RGB samples of camera sensors in the 2-Lane scenario.}
                \label{rgbsample_2Lane}
        \end{subfigure}%
        \begin{subfigure}[t]{0.45\textwidth}
        \centering
               \captionsetup{justification=raggedright,singlelinecheck=false}
               \includegraphics[width=\linewidth]{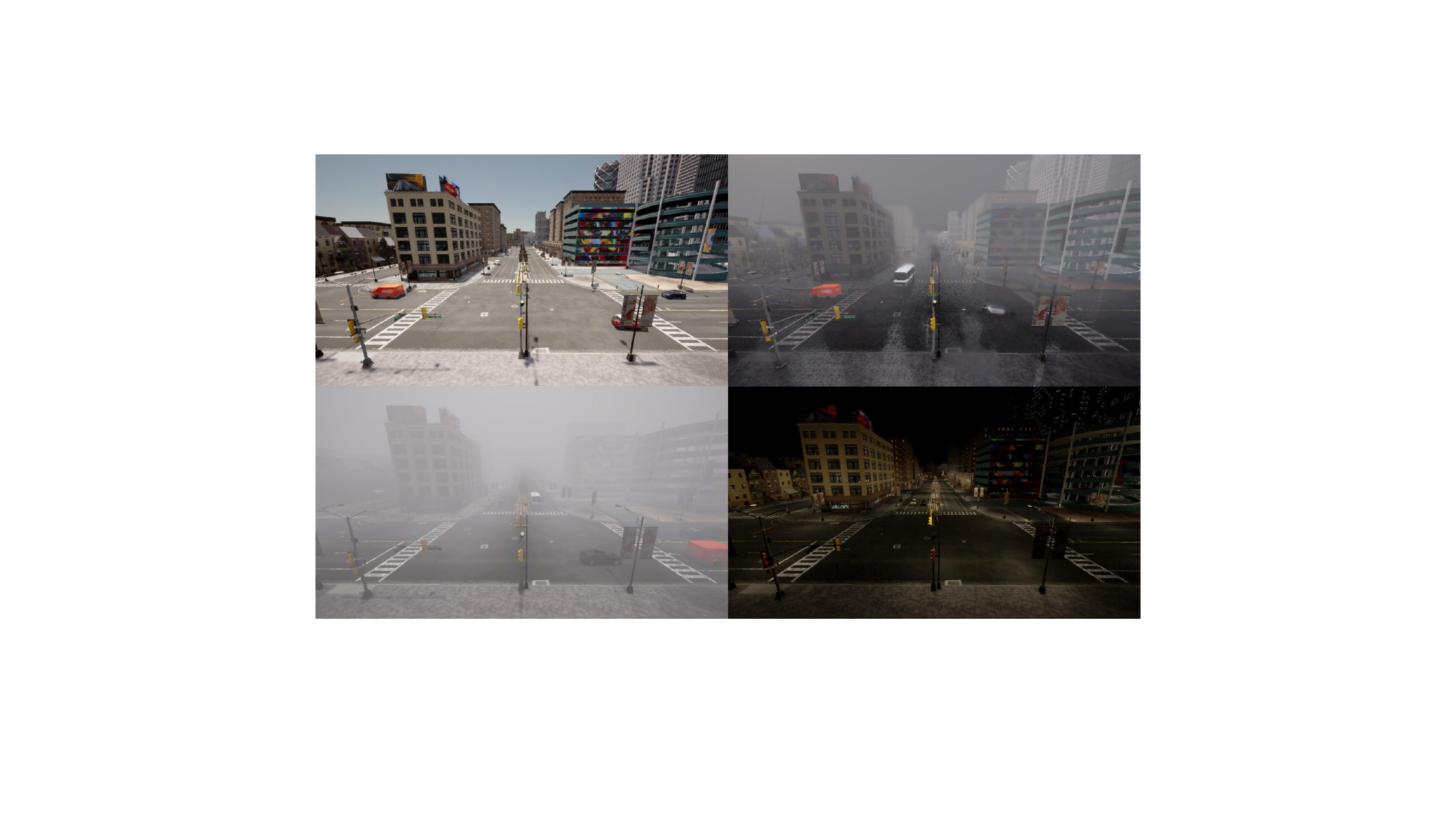}
                \caption{RGB samples of camera sensors in the 3-Lane scenario.}
                \label{rgbsample_3Lane}
        \end{subfigure}%
        \caption{RGB samples of camera sensors by episode type.}\label{fig:rgb_samples}
\vspace{-.175in}
\end{figure*}
\begin{figure*}[!t]
        \centering
        \begin{subfigure}[t]{0.29\textwidth}
        \centering
               \captionsetup{justification=raggedright,singlelinecheck=false}
                \includegraphics[width=\linewidth]{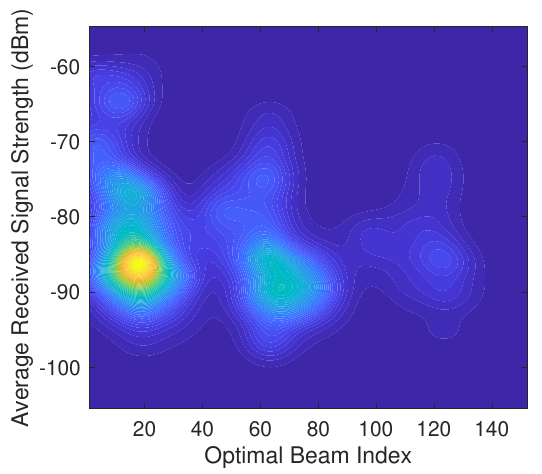}
                \caption{Distribution of 2-Lane Scenario.}
                \label{distribution_2Lane}
        \end{subfigure}%
        \begin{subfigure}[t]{0.29\textwidth}
        \centering
               \captionsetup{justification=raggedright,singlelinecheck=false}
               \includegraphics[width=\linewidth]{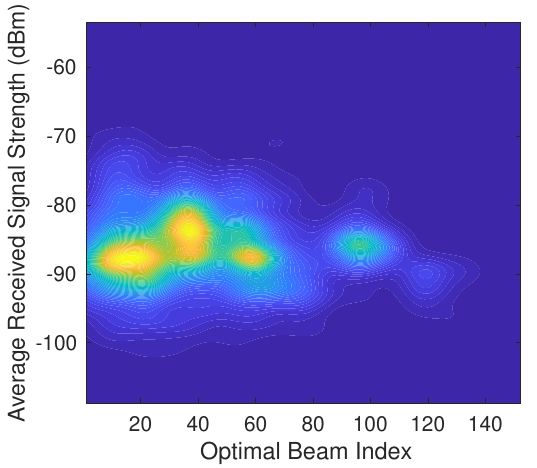}
                \caption{Distribution of 3-Lane Scenario.}
                \label{distribution_3Lane}
        \end{subfigure}%
        \begin{subfigure}[t]{0.326\textwidth}
        \centering
             \captionsetup{justification=raggedright,singlelinecheck=false}
                \includegraphics[width=\linewidth]{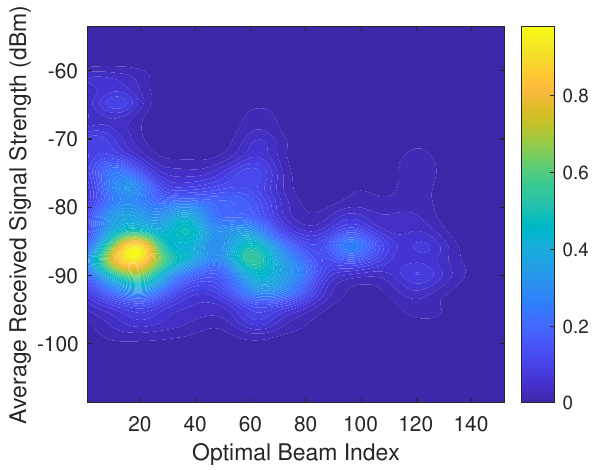}
                \caption{Distribution of all dataset.}
                \label{distribution_all}
        \end{subfigure}%
        \caption{Analysis of distributions of the generated dataset.}\label{distributions}
\vspace{-.175in}
\end{figure*}

\begin{figure}[t]
    \centering
    \includegraphics[width=0.75\columnwidth]{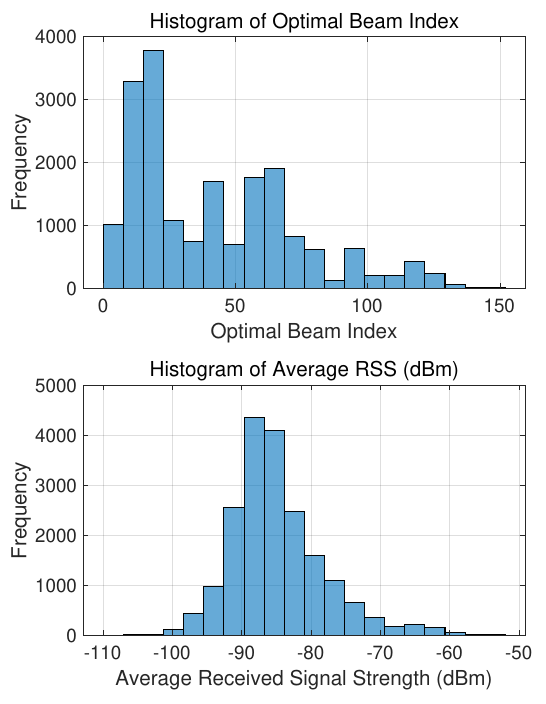}
    \caption{Distribution analysis of the generated dataset.}
    \label{fig:hist_all_dataset}
\vspace{-.175in}
\end{figure}

\color{black}
\section{Simulation Results and Analysis}
\label{exp_results}

\textcolor{black}{A simulation-based dataset was generated using the `Town 10' map in CARLA, designed to resemble an urban environment with up to 40 vehicles moving along 2-Lane or 3-Lane roads.  Each scenario (2-Lane or 3-Lane) comprises 50 episodes, spanning a maximum of 200 time steps sampled at $100$ ms intervals. This setup produces $18,823$ time samples ($9,073$ in the 2-Lane scenario and $9,750$ in the 3-Lane scenario).} To introduce additional variation, the simulation adjusts the time of day and weather: $30$ episodes occur at noon, $10$ at night, and the remaining $10$ episodes feature rain or fog in equal proportions. Figure~\ref{fig:rgb_samples} provides example RGB frames that illustrate these conditions. 

We train the networks in $70\%$ of the generated samples, evaluated in $15\%$, and tested the remaining $15\%$. All results are reported in the test set. We implemented our models in PyTorch. The teacher network was trained for $50$ epochs with a batch size of $64$ and an initial learning rate of $5\times 10^{-4}$ with the learning rate decayed by an absolute amount of $5\times 10^{-6}$ per epoch. A learning rate decay starts at epoch $15$, and a restart interval is applied every $10$ epochs. The student was trained for 50 epochs in the same settings. We set the loss weight of the distillation $\alpha=0.5$. Within this virtual environment, a sensing-aided beamformer is used to serve the vehicles, and 152 distinct beam patterns are defined to cover both single-beam and multi-beam configurations of up to three beams. Each beam pattern spans an elevation angle of \(70^\circ\) and an azimuth angle of \(180^\circ\). At each time step, the beam pattern that maximizes the average received signal strength (RSS) across multiple vehicles is designated as optimal. In order to assess the performance of our beam prediction model from multiple perspectives, we employ three distinct metrics: Top-$k$ accuracy, mean received signal strength (RSS) and the mean percentile rank (MPR).

\noindent
\emph{1) Top-$k$ Accuracy}: 
This metric checks whether the optimal beam $y$ appears among the top candidates $k$ predicted according to the model’s probability distribution $\boldsymbol{p}$. $\mathrm{Top}_k(\boldsymbol{p})$ is the set of beam indices with the highest probabilities $k$ in $\boldsymbol{p}$. Formally, we can express Top-$k$ Accuracy as follows:

\begin{equation}
\label{eq:topk}
\mathrm{Top\text{-}k\ Accuracy}
=
\frac{1}{N}\sum_{n=1}^{N}
\mathbb{I}\Bigl(y^{(n)} \,\in\, \mathrm{Top}_k\bigl(\boldsymbol{p}^{(n)}\bigr)\Bigr),
\end{equation}
where $N$ is the total number of test samples, and $\mathbb{I}(\cdot)$ is an indicator function that returns $1$ if the condition is true and $0$ otherwise. A higher Top-$k$ value indicates that the model’s probability distribution consistently assigns high scores to the correct beam within its top predictions.

\noindent
\emph{2) Mean RSS}:
Once the model selects $\hat{y}$ as the predicted beam index, the system applies the corresponding beamformer, resulting in a received signal strength $S(\hat{y})$. We define the mean RSS as the average RSS achieved over the test set as follows:

\begin{equation}
\label{eq:mean_rss}
\overline{S}
=
\frac{1}{N}
\sum_{n=1}^{N}
S\Bigl(\,\hat{y}^{(n)}\Bigr).
\end{equation}
A higher value of $\overline{S}$ implies that the beam prediction model more often aligns with beam patterns maximizing signal strength across diverse scenarios.

\vspace{1mm}
\noindent
\emph{3) Mean Percentile Rank (MPR)}:
The percentage rank offers another view of the prediction quality by examining the rank of the predicted beam among all $B$ beams in terms of the actual RSS performance. We define the rank of $\hat{y}$ as:

\begin{equation}
\mathrm{rank}\bigl(\hat{y}\bigr)
=
1
+ \sum_{b=1}^{B}
\mathbb{I}\Bigl(S(b) > S(\hat{y})\Bigr),
\end{equation}
so that a rank of $1$ indicates the best (highest RSS) beam. We convert this rank to a percentile by:

\begin{equation}
\mathrm{Percentile}(\hat{y})
=
\frac{B - \mathrm{rank}(\hat{y}) + 1}{B},
\end{equation}
which ranges from $0$ to $1$. The MPR over the test set can be written as:

\begin{equation}
\label{eq:mpr}
\mathrm{MPR}
=
\frac{1}{N}
\sum_{n=1}^{N}
\mathrm{Percentile}\Bigl(\,\hat{y}^{(n)}\Bigr).
\end{equation}
MPR essentially measures how close the chosen beam is to optimal in terms of percentile. An MPR of $100\%$ means the top-ranked beam was always chosen, whereas lower values indicate the prediction often fell short of the best beam. This measure is particularly useful when analyzing how close the prediction is to the truly optimal beam in a ranked sense, rather than a strict “correct vs. incorrect” classification.

\subsection{Dataset Analysis}
As shown in Fig.~\ref{distributions}, the most frequently chosen beam indices tend to be single-beam solutions, especially in the 2-Lane scenario, where line of sight or partial visibility to one dominant path is more likely. \textcolor{black}{Although 3-Lane roads lead to more multibeam usage, single-beam choices remain prevalent}. This indicates that in many time steps, one beam is sufficient to cover the dominant paths and using additional beams (while possible) does not substantially increase RSS. Hence, the optimal solution skews toward single beams. Multibeam patterns are only chosen when vehicles or obstructions create multiple equally strong paths that a single beam cannot cover. The distribution of optimal beams in these experiments illustrates that multi-beam patterns, notably those with three closely spaced beams, are seldom chosen as they do not substantially improve average RSS compared to strong single-beam alignments. 

Figure~\ref{fig:hist_all_dataset} further shows a skew in the usage of beam indexes: certain indices appear far more frequently than others, indicating an inherent label imbalance. This imbalance reflects practical conditions where one or two strong angles can dominate the channel environment, rendering most other beam patterns suboptimal. This outcome highlights a key training consideration for learning-based beam prediction since models must handle real-world datasets where some classes (i.e., certain beam indices) are heavily favored.

\begin{table*}[!htp]
    \centering
    \caption{Learning results of cross-modal relational knowledge distillation between models with the same training data (2-Lane, 3-Lane, All).}
    \begin{tabular}{ccccccccc}
        \toprule
        & & & \multicolumn{6}{c}{\textbf{Scenarios}} \\
        \cmidrule(l){4-9}
        \multicolumn{3}{c}{\textbf{Methods}} & \multicolumn{2}{c}{\textbf{2-Lane}} & \multicolumn{2}{c}{\textbf{3-Lane}} & \multicolumn{2}{c}{\textbf{All}} \\
        \cmidrule(lr){1-3} \cmidrule(lr){4-5} \cmidrule(lr){6-7} \cmidrule(l){8-9}
        & \textbf{Latent} & \textbf{Output} & \textbf{MPR (\%)} & \textbf{RSS (dBm)} 
        & \textbf{MPR (\%)} & \textbf{RSS (dBm)} 
        & \textbf{MPR (\%)} & \textbf{RSS (dBm)}\\
        \midrule
        \multicolumn{3}{c}{Teacher (105M)} &  92.172 & -77.753 & 84.362 & -84.148 & 88.410 & -81.276 \\
        \multicolumn{3}{c}{WithoutKD (13M)} & 84.120 & -82.407 & 78.288 & -86.803 & 80.644 & -84.827 \\
        \midrule
         & KLdiv& - & 83.776 & -82.561 & 78.293 & -86.615 & 82.019 & -84.538 \\
        KD  & - & KLdiv & 84.426 & -82.322 & 79.503 & -86.551 & 82.783 & -84.412 \\
        & KLdiv & KLdiv & 86.329 & -81.399 & 79.860 & -86.155 & 82.336 & -84.209 \\
        \midrule
         & manifold & - & 86.998 & -81.266 & 79.525 & -86.395 & 83.551 & -84.006 \\
         & - & manifold & 85.844 & -81.627 & 79.707 & -86.323 & 83.463 & -84.158 \\
        RKD & manifold & manifold & 87.210 & -81.084 &  80.101 & -85.708 & 83.658 & -83.603 \\
         & - & beamStr & 87.185 & -80.690 & 80.902 & -85.624 & 83.276 &  -83.979 \\
         & manifold & beamStr & \textbf{87.502} & \textbf{-80.610} &  \textbf{81.116} & \textbf{-85.534} & \textbf{83.972} & \textbf{-83.526} \\
        \bottomrule
    \end{tabular}
    \label{result_table1}
\end{table*}

\begin{figure*}[t]
    \centering
    \includegraphics[width=\textwidth]{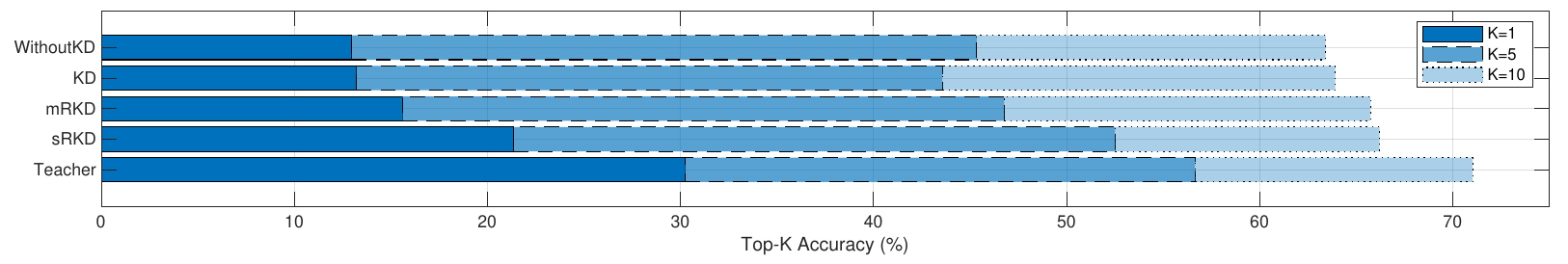}
    \caption{Comparative analysis of Top-K (1,5,10) prediction accuracy across models trained with all datasets.}
    \label{result_topk}
\end{figure*}

\begin{table}[t]
\centering
\caption{Comparison of MPR (\%) with and without RKD across different student model sizes.}
\label{tab:network_size}
\begin{tabular}{lccc}
\toprule
\textbf{Model Size} & \textbf{Without KD} & \textbf{With RKD} & \textbf{Gain} \\
\midrule
Student-S (6M)  & 83.27 & 86.77 & \textbf{+3.50} \\
Student-M (13M) & 84.20 & 87.50 & \textbf{+3.30} \\
Student-L (34M) & 86.36 & 88.45 & \textbf{+2.09} \\
\bottomrule
\end{tabular}
\end{table}

\begin{table}[t]
\centering
\caption{Comparison of model complexity and inference latency across platforms.}
\label{tab:latency}
\begin{tabular}{lcccccc}
\toprule
\textbf{Model} & \textbf{\#Params} & \textbf{FLOPs} & \textbf{GPU} & \textbf{CPU1} & \textbf{CPU2} \\
 & & & (ms) & (ms) & (ms) \\
\midrule
Teacher         & 105.6M & 20.3G & 24.52 & 30.03 & 946.18 \\
Student-L & 34.2M  & 34.2M & 0.66  & 0.99  & 7.99   \\
Student-M & 13.6M  & 13.6M & 0.73  & 1.10  & 4.93   \\
Student-S & 6.8M   & 6.8M  & 0.59  & 1.04  & 2.21   \\
\bottomrule
\end{tabular}
\vspace{-.175in}
\end{table}

\begin{table*}[!t]
\centering
\caption{Comparison of beam prediction results under domain-same (2-Lane) and domain-shift (3-Lane to 2-Lane) scenarios.}
\label{tab:domain_shift}
\begin{tabular}{l cc cc cc cc}
\toprule
& \multicolumn{2}{c}{\textbf{Domain Same (2-Lane)}} & \multicolumn{6}{c}{\textbf{Domain Shift (3-Lane to 2-Lane)}} \\
\cmidrule(lr){2-3} \cmidrule(lr){4-9}
& \multicolumn{2}{c}{\textbf{2-Lane}} & \multicolumn{2}{c}{\textbf{2-Lane}} 
& \multicolumn{2}{c}{\textbf{3-Lane}} & \multicolumn{2}{c}{\textbf{Average}} \\
\cmidrule(lr){2-3} \cmidrule(lr){4-5} \cmidrule(lr){6-7} \cmidrule(lr){8-9}
\textbf{Methods} 
& \textbf{MPR (\%)} & \textbf{RSS (dBm)} 
& \textbf{MPR (\%)} & \textbf{RSS (dBm)} 
& \textbf{MPR (\%)} & \textbf{RSS (dBm)} 
& \textbf{MPR (\%)} & \textbf{RSS (dBm)} \\
\midrule
Teacher & 92.172 & -77.753 & 85.496 & -81.782 & 84.362 & -84.148 & 84.929 & -82.965 \\
WithoutKD & 84.199 & -82.406 & 53.307 & -94.567 & 78.287 & -86.802 & 65.797 & -90.685 \\
sRKD    & \textbf{87.502} & \textbf{-80.610} & 54.781 & -94.088 & \textbf{81.115} & \textbf{-85.625} & \textbf{67.802} & \textbf{-90.177} \\
mRKD    & 85.329 & -81.889 & \textbf{55.037} & \textbf{-93.998} & 80.101 & -85.708 & 66.838 & -90.383 \\
KD     & 84.849 & -82.121 & 54.901 & -94.007 & 79.860 & -86.155 & 66.348 & -90.476 \\
\bottomrule
\end{tabular}
\vspace{-.175in}
\end{table*}

\subsection{Beam Prediction Accuracy}
Table~\ref{result_table1} summarizes the performance of cross-modal knowledge distillation for beam prediction. The \emph{Teacher} network, a transformer-based model trained on multimodal sensing input (LiDAR, radar, GPS, and RGB), achieves the highest-rank accuracies of $92.17\%$, $84.36\%$, and $88.41\%$ in the scenarios tested. The 3-Lane scenario, having more complex multipath conditions, naturally results in lower accuracy across all models (\emph{Teacher}'s MPR drops by approximately 8 points compared to the 2-Lane scenario), highlighting the added difficulty. The \emph{WithoutKD} model is a radar-only MLP network without any distillation. The \emph{WithoutKD} model reaches $84.12\%$, $78.29\%$, and $82.02\%$, demonstrating a lower performance than \emph{Teacher}. This gap underscores the value of multimodal sensor data in guiding beam selection. We apply distillation at two levels of the network: the latent layer (the latent features, e.g. after the encoder or Transformer module) and the output layer (the final output logits before the softmax). \textcolor{black}{Among the standard KD baselines (using KL divergence), \emph{KD (KLdiv+KLdiv)} reaches up to $86.32\%$, $79.86\%$, and $82.34\%$ but still underperforms compared to the RKD approaches. In the RKD section, various combinations of latent- and output-level relation transfer are explored: \emph{RKD (manifold+beamStr)} achieves the highest student performance, with an MPR of $87.50\%$ on the 2-Lane and $83.97\%$ in the overall dataset. This configuration also yields the best RSS values (e.g. $-80.61$ dBm in 2-Lane), confirming its superior beam selection accuracy. The improvement of RKD over traditional KD methods illustrates the importance of transferring structured relationships, especially those based on beamforming-specific steering vector similarities (beamStr).} Overall, these results indicate that even a radar-only student network can achieve high accuracy if it benefits from cross-modal distillation of teacher feature relationships. In practical terms, this finding suggests that a small, cost-effective network, limited to radar sensing for real-time inference, can still approximate the performance of a more complex multimodal teacher.

\textcolor{black}{Fig.~\ref{result_topk} illustrates the Top-K prediction accuracy ($K=1,5,10$) across five different models trained using all datasets. The baseline \emph{WithoutKD} (radar-only MLP without distillation) shows the lowest performance across all K values, reflecting the limitations of lightweight monomodal learning. The \emph{KD} model refers to conventional knowledge distillation using KL divergence at both the latent and output layers (\emph{KD (KLdiv+KLdiv)}). The \emph{KD} improves the accuracy of Top-1 and Top-10 compared to baseline, but remains less effective in capturing spatial beam relationships. The \emph{mRKD} model represents relational knowledge distillation applied to both latent features (\emph{RKD (manifold+manifold)}, while \emph{sRKD} applies RKD jointly on latent and output-level beam steering representations (\emph{RKD (manifold+beamStr)}). Among the two, \emph{sRKD} achieves superior performance, especially in Top-1 and Top-5 metrics, highlighting the benefit of beam-aware similarity modeling in the output space. The \emph{Teacher} model, trained on full multimodal input with transformer architecture, achieves the highest overall accuracy, serving as an upper bound. However, \emph{sRKD} narrows the performance gap in Top-10 accuracy while maintaining a lightweight radar-only architecture, demonstrating the effectiveness of beam-aware relational distillation under resource constraints.}

Table~\ref{tab:network_size} compares the performance of RKD in different model sizes for the student network. The results indicate that as the MLP architecture of the student grows, the overall precision increases both for the baseline approach (\emph{WithoutKD}) and the relational distillation approach (\emph{RKD}). However, the relative improvement over KD is more pronounced in smaller student networks. In other words, while \emph{RKD} achieves greater absolute accuracy with larger models, the performance gap between \emph{WithoutKD} and \emph{RKD} is notably greater when the student model is small. This suggests that knowledge distillation can be especially advantageous in resource-constrained scenarios, where the student network is required to maintain a low parameter count while striving to approximate the performance of a significantly more complex multimodal teacher. Consequently, compact radar-only models can benefit substantially from cross-modal distillation, realizing stronger beam prediction capabilities without incurring the computational overhead that larger networks demand.

\color{black}
\subsection{Resource Efficiency Analysis}
Table~\ref{tab:latency} presents a comparison of model size, FLOPs, and inference latency for teacher and three student variants on three hardware platforms: a GPU (NVIDIA RTX 4070), a desktop CPU (AMD Ryzen 7 7800X3D @ 4.2GHz) and a mobile CPU (Intel i7-10510U @ 1.8GHz). While the teacher model, based on a Transformer architecture, incurs high computational cost (20.3G FLOPs) and latency up to 946.18ms on mobile CPUs, the student models are designed with lightweight MLPs. As a result, their FLOPs closely match their parameter counts (e.g. Student-S: 6.8M parameters $\approx$ 6.8M FLOPs), enabling significant speed up during inference. In particular, the smallest student model achieves 0.59ms on GPU and 1.04ms on CPU1, satisfying the sub-1ms latency target for 5G/6G beamforming, especially under ultra-reliable low-latency communication (URLLC) requirements. Moreover, given that standard 5G NR beam sweeping takes approximately 20ms, our method provides over 30× faster inference with competitive accuracy. This indicates the practical deployability of our method for real-time mmWave beam prediction on edge and mobile devices, without compromising performance.

\color{black}
\subsection{Limitations under Domain Shift and Future Directions}
In addition to evaluating performance under matched training and testing domains, we also investigate a domain-shift scenario to assess how well knowledge distillation withstands different environmental conditions. Specifically, the teacher network is trained on a 2-Lane dataset and then used to distill knowledge into a student network trained on a 3-Lane dataset. After this cross-domain distillation, the student model is evaluated in both the 2-Lane and 3-Lane test sets. Table~\ref{tab:domain_shift} presents the beam prediction performance in both domain-matched (2-Lane) and domain-shifted scenarios, where the student model is trained on 3-Lane data and tested on 2-Lane and 3-Lane environments. The teacher model, trained on 2-Lane data, is also evaluated on both test sets for comparison. The Domain Shift setting specifically examines how well the student generalizes when trained and tested in different environments. The results show a noticeable drop in student performance (of up to $32.72\%$) under domain shift, especially in the 2-Lane test set, indicating limited generalization capacity. This performance degradation is more severe for the radar-only student due to its lower model capacity and restricted sensor input. In contrast, the multimodal teacher, with richer inputs and a transformer-based architecture, exhibits more stable performance across domains. These findings highlight the vulnerability of compact single-modal models to domain mismatch and underscore the importance of cross-domain robustness in real-world deployments.

Although domain-shifted models show noticeable performance degradation compared to domain-matched training, particularly in radar-only settings, these results underscore the practical challenge of generalizing to unseen environments. In real-world deployments, sensing-aided beamformers are unlikely to encounter the same environmental conditions as those used during training, making robustness to domain shift a critical requirement. Our findings reveal that compact radar-only student models are particularly sensitive to such shifts, likely due to their limited input diversity and lower representational capacity. This highlights the need for more effective domain generalization strategies, such as domain-invariant feature learning, data augmentation across diverse scenarios, or lightweight fine-tuning mechanisms. In future work, we aim to explore these approaches in depth and incorporate domain-shift robustness as a core evaluation criterion, with the goal of enabling resource-constrained student models to maintain reliable performance across heterogeneous deployment environments.

\section{Conclusion}
\label{conclusion}
In this paper, we have developed a novel CRKD framework for efficient mmWave beam prediction, in which a multimodal teacher model (LiDAR, radar, GPS, and RGB) transfers relational knowledge to a radar-only student. To achieve realistic evaluations, we have integrated CARLA-based sensor data generation with MATLAB-based mmWave channel modeling, creating a comprehensive simulation environment. Experimental results in 2-Lane and 3-Lane road scenarios confirmed that a compact student model with radar can only approach the performance of a transformer-based teacher rich in sensors. This indicates that real-world applications remain viable even with fewer sensor modalities and a smaller network structure. In future work, we plan to incorporate domain adaptation techniques into the CRKD process to further enhance the generalization of the student model under changing conditions. We will also integrate additional sensors and more complex mobility patterns into our simulation framework, with the aim of maintaining high beam prediction accuracy in resource-constrained deployment scenarios and across diverse real-world environments.

\ifCLASSOPTIONcaptionsoff
  \newpage
\fi
\bibliographystyle{IEEEtran}
\bibliography{ref}
\end{document}